\documentclass{jfm}

\usepackage{graphicx}
\usepackage{xcolor}
\usepackage{natbib}
\usepackage{amsmath}
\usepackage{amssymb}
\usepackage{ar}

\newcommand{\ie}{i.e.,\ }
\newcommand{\eg}{e.g.,\ }

\newcommand{\secref}[1]{\mbox{section \ref{#1}}}

\newcommand{\tabref}[1]{\mbox{table \ref{#1}}}
\newcommand{\equrefC}[1]{\mbox{Equation (\ref{#1})}}
\newcommand{\equref}[1]{\mbox{equation (\ref{#1})}}

\newcommand{\figrefC}[2][]{\mbox{Figure \ref{#2}(#1)}}
\newcommand{\figref}[2][]{\mbox{figure \ref{#2}(#1)}}
\newcommand{\figrefSC}[1]{\mbox{Figure \ref{#1}}}
\newcommand{\figrefS}[1]{\mbox{figure \ref{#1}}}

\title[Droplets in homogeneous shear turbulence]
{Droplets in homogeneous shear turbulence}

\author[Rosti, Ge, Jain, Dodd and Brandt]{Marco E. Rosti$^1$\thanks{Email address for correspondence: merosti@mech.kth.se}, Zhouyang Ge$^1$, Suhas S. Jain$^2$, Michael S. Dodd$^2$ and Luca Brandt$^{1,3}$}

\affiliation{$^1$ Linn\'{e} Flow Centre and SeRC, KTH Mechanics, Stockholm, Sweden \\$^2$ Center for Turbulence Research, Stanford University, USA \\$^3$ Department of Energy and Process Engineering, Norwegian University of Science and Technology (NTNU), Norway}

\begin{document}
\maketitle

\begin{abstract}
We simulate the flow of two immiscible and incompressible fluids separated by an interface in a homogeneous turbulent shear flow at a shear Reynolds number equal to $15200$. The viscosity and density of the two fluids are equal, and various surface tensions and initial droplet diameters are considered in the present study. We show that the two-phase flow reaches a statistically stationary turbulent state sustained by a non-zero mean turbulent production rate due to the presence of the mean shear. Compared to single-phase flow, we find that the resulting steady state conditions exhibit reduced Taylor microscale Reynolds numbers owing to the presence of the dispersed phase, which acts as a sink of turbulent kinetic energy for the carrier fluid. At steady state, the mean power of surface tension is zero and the turbulent production rate is in balance with the turbulent dissipation rate, with their values being larger than in the reference single-phase case. The interface modifies the energy spectrum by introducing energy at small-scales, with the difference from the single-phase case reducing as the Weber number increases. This is caused by both the number of droplets in the domain and the total surface area increasing monotonically with the Weber number. This reflects also in the droplets size distribution which changes with the Weber number, with the peak of the distribution moving to smaller sizes as the Weber number increases. We show that the Hinze estimate for the maximum droplet size, obtained considering breakup in homogeneous isotropic turbulence, provides an excellent estimate notwithstanding the action of significant coalescence and the presence of a mean shear.
\end{abstract}

\section{Introduction} \label{sec:introduction}
The understanding of liquid-liquid emulsions is important in many industrial processes \eg hydrocarbon separation, suspension crystallization, and emulsion polymerization. These flows are characterized by density and viscosity ratios on the order of unity (\eg water and oil mixtures) and a source of agitation (\eg an impeller) that creates a turbulent two-phase mixture consisting of a dispersed phase of droplets and a continuous phase. The resulting turbulence in the carrier phase is altered directly by the droplet feedback on the surrounding fluid and indirectly by droplet-droplet interactions. Many aspects of the complex interaction of the dispersed phase with the continuous phase are not well understood. In particular, there are questions related to the topological changes and to the role of the surface tension of the dispersed phase, the stationarity of the turbulent statistics, and the kinetic energy budget.

Liquid-liquid emulsions have been the subject of numerous experimental \citep{berkman_calabrese_1988a, pacek_man_nienow_1998a, lovick_mouza_paras_lye_angeli_2005a} and computational studies \citep{perlekar_biferale_sbragaglia_srivastava_toschi_2012a, skartlien_sollum_schumann_2013a, komrakova_eskin_derksen_2015a, scarbolo_bianco_soldati_2015a, dodd_ferrante_2016a}. The computational studies can be broadly categorized as forced homogeneous isotropic turbulence \citep{perlekar_biferale_sbragaglia_srivastava_toschi_2012a, skartlien_sollum_schumann_2013a, komrakova_eskin_derksen_2015a}, decaying homogeneous isotropic turbulence \citep{dodd_ferrante_2016a} and turbulent wall flows \citep{scarbolo_bianco_soldati_2015a}. Forced homogeneous isotropic turbulence has the advantage of producing a statistically homogeneous and isotropic flow field that, in time, can reach a statistically stationary state. However, in forced homogeneous isotropic turbulence, the turbulent kinetic energy must be induced artificially via a forcing term in the Navier--Stokes equations. This is in contrast to a natural forcing mechanism that produces turbulent kinetic energy from finite Reynolds stresses interacting with a mean velocity gradient. While forcing homogeneous isotropic turbulence may be appropriate for studying the droplet size distributions, it has been argued that artificial forcing is inappropriate for studying two-way coupling effects \citep{elghobashi_2019a}. Therefore, for studying the turbulent kinetic energy budget, either decaying isotropic turbulence or turbulent shear flow might be preferable.

In decaying isotropic turbulence, it was shown that the presence of finite-size droplets always enhances the decay rate of the turbulent kinetic energy \citep{dodd_ferrante_2016a}. Also, the deformation, breakup, or coalescence of the droplets introduces an additional term to the turbulent kinetic energy equation - the power of the surface tension - termed $\Psi_\sigma$ by \cite{dodd_ferrante_2016a}, which describes the rate of change of the interfacial energy, balancing the kinetic energy transfer between the external fluid and the flow inside the droplets. Correct identification of these pathways for the turbulent kinetic energy exchange is fundamental to understand the turbulence modulation by the droplets and then to model it.

Building upon previous studies, we consider finite-size bubbles/droplets of Taylor length scale in homogeneous shear turbulence \citep{tavoularis_corrsin_1981b, tavoularis_corrsin_1981a, pumir_1996a, mashayek_1998a, sekimoto_dong_jimenez_2016a}. Homogeneous shear turbulence flow is conceivably the simplest case in which the flow remains statistically homogeneous in all spatial directions. Moreover, compared to forced isotropic turbulence, it has a natural energy production mechanism via a mean velocity gradient. We note that ideal homogeneous shear turbulence is self-similar, implying an unbounded energy growth within infinite domains \citep{sukheswalla_vaithianathan_collins_2013a}. This condition limits any numerical simulations to relatively short times, concerning only the initial shearing of isotropic turbulence \citep{rogers_moin_1987a, lee_kim_moin_1990a, sukheswalla_vaithianathan_collins_2013a}. However, as demonstrated by \cite{pumir_1996a} and \cite{sekimoto_dong_jimenez_2016a} in single-phase flow, the finite computational box introduces a large-scale confinement effect similar to that enforced by a wall; thus, a meaningful statistically stationary state can be reached over long periods, termed statistically stationary homogeneous shear turbulence (SS--HST). In particular, \cite{sekimoto_dong_jimenez_2016a} showed that long-term simulations of HST are ``minimal" in the sense of containing on average only a few large-scale structures: all the one-point statistics agree well with those of the logarithmic layer in turbulent channel flows, particularly when scaled with the friction velocity derived from the measured Reynolds stresses. The same holds for the wall-parallel spectra of the wall-normal velocity. The authors concluded that the similarities between the steady state homogeneous shear turbulence and other shear flows, particularly with the logarithmic layer of wall turbulence, make it a promising system to study shear turbulence in general. These observations, combined with the insights recently gained in the droplet-turbulence interaction in decaying homogeneous isotropic turbulence, motivate us to further investigate turbulence modulation due to droplets/bubbles in steady state homogeneous shear turbulence.

In this paper, we present DNS of an emulsion created by droplets dispersed in homogeneous shear turbulence. By changing the initial size of the dispersed phase and the Weber number, we aim to answer the following questions:
\begin{enumerate}
\item[(a)] ~Can a statistically stationary state be reached when the dispersed phase actively undergoes breakup and coalescence in homogeneous shear turbulence?
\item[(b)] ~If so, what determines the steady-state size distribution of the dispersed phase?
\item[(c)] ~How does the dispersed phase change the turbulent kinetic energy budget?
\end{enumerate}
Homogeneous shear turbulence shares many similarities with other shear flows, including turbulent wall flows \citep{sekimoto_dong_jimenez_2016a}; therefore, by answering these questions, we expect to improve our understanding of the droplet-turbulence interaction and, hopefully, help future modelers gain intuition about more complex conditions. 

To capture the complex phenomena accurately in a direct numerical simulation of turbulent two-phase flow, we need a numerical method that is reliable and possess the following properties: (i) discrete mass, momentum and kinetic energy conservation, (ii) ability to handle large jumps in density, (iii) ability to handle complex topologies and separation of scales, and (iv) accurate surface tension implementation \citep{mirjalili_jain_dodd_2017a}. In the present work, we choose to use an algebraic volume of fluid method known as THINC (tangent of hyperbola for interface capturing) method, which is a sharp-interface method. This method is relatively new and has been demonstrated to be as accurate and also cost effective compared to the well known geometric volume of fluid methods in canonical test cases \citep{xie_ii_xiao_2014a}, which makes it a good alternative. However, \citet{mirjalili_jain_dodd_2017a} indicate that large-scale realistic simulations of turbulent two-phase flows using THINC methods are still lacking in the literature and are crucial to fully evaluate the capabilities of these methods \citep[see][for the use of the THINC method for low Reynolds number flows]{rosti_de-vita_brandt_2019a}. Hence we choose to use this method in the current study, which will serve as an evaluation of the robustness of THINC methods for complex realistic simulations. 

This paper is organized as follows. In \secref{sec:formulation}, we first discuss the flow configuration and the governing equations and then present the numerical methodology used. The results on the fully developed two-phase homogeneous shear turbulent flow are presented in \secref{sec:result}, where we answer the questions discussed above based on our observations. In particular, we first show how the turbulent flow is modified by the droplets and how the droplets evolve in the turbulent flow, and then explain how these modifications occur by studying the turbulent kinetic energy balance in the two-phase flow. Finally, all the main findings and conclusions are summarized in \secref{sec:conclusion}.

\section{Methodology} \label{sec:formulation}
\subsection{Governing equations and numerical methods}
\begin{figure}
	\centering
	\includegraphics[width=0.8\textwidth]{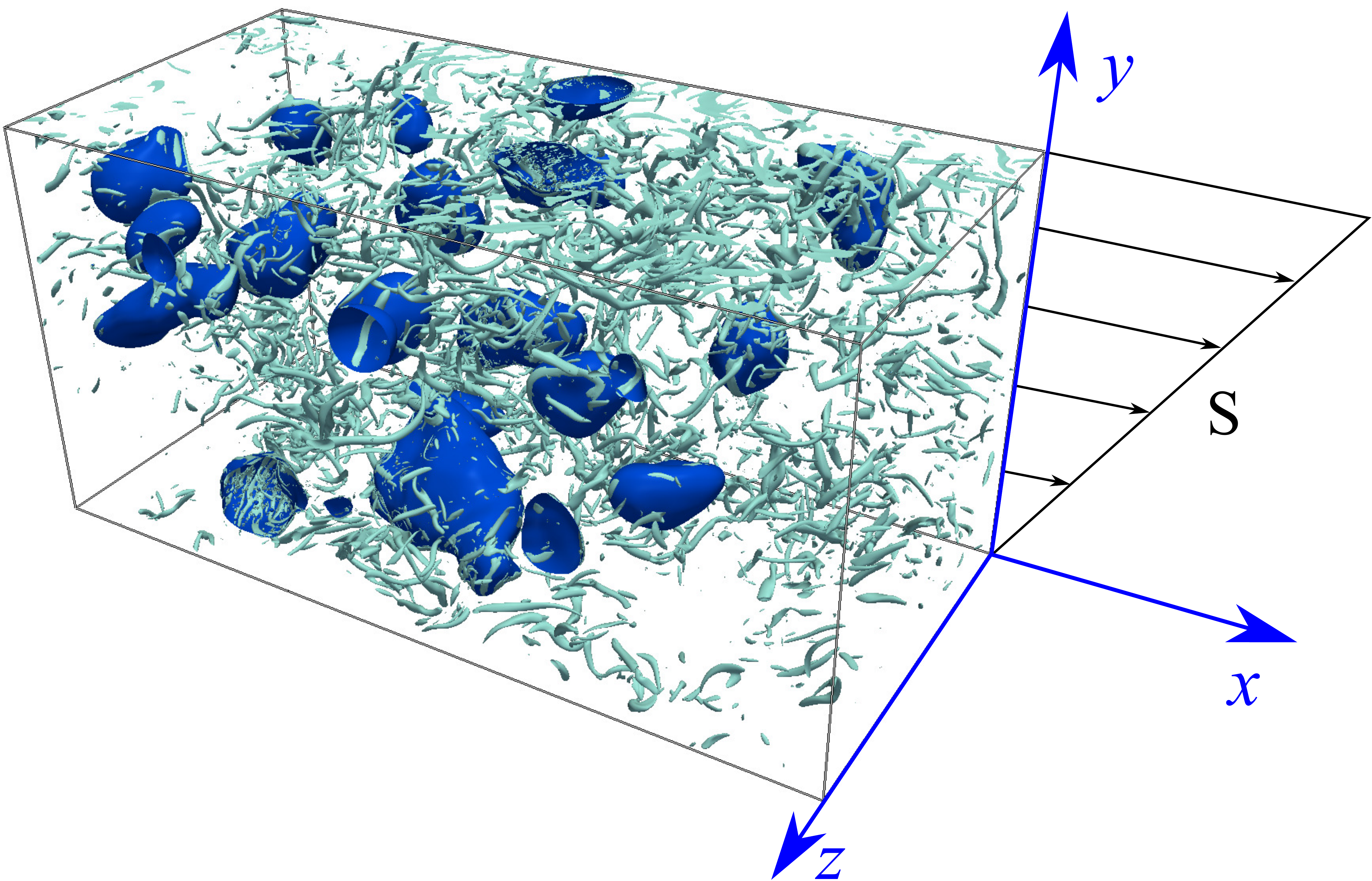}
	\caption{Sketch of the computational domain and of the Cartesian coordinate system. The visualization pertains the flow at $Re_z\approx 15000$ with $5\%$ volume fraction of the dispersed phase at $We_\lambda \approx 0.75$. The blue color is used to depict the surface of the droplets.}
	\label{fig:sketch}
\end{figure}
We consider the flow of two immiscible incompressible fluids in a periodic box subject to a uniform mean shear $\mathcal{S}$. \figrefSC{fig:sketch} shows a sketch of the geometry and the Cartesian coordinate system, where $x$, $y$, and $z$ ($x_1$, $x_2$, and $x_3$) denote the streamwise, shear, and spanwise coordinates, and $u$, $v$, and $w$ ($u_1$, $u_2$, and $u_3$) denote the respective components of the velocity field. Standard periodic conditions are applied in $x$ and $z$, and a shear-periodic boundary condition is enforced in $y$, \ie
\begin{align}
u_i \left( x_1+L_x, x_2, x_3 \right) &= u_i \left( x_1, x_2, x_3 \right), \\
u_i \left( x_1, x_2+L_y, x_3 \right) &= u_i \left( x_1-\mathcal{S}tL_2, x_2, x_3 \right), \\
u_i \left( x_1, x_2, x_3+L_z \right) &= u_i \left( x_1, x_2, x_3 \right). 
\end{align}

The total velocity field $u_i$ can be decomposed for convenience into the sum of a mean component $\langle u_i \rangle_{xz}$ generated by the imposed shear $\mathcal{S}$, \ie $\langle u_i \rangle_{xz}=\mathcal{S}x_2 \delta_{1i}$ where $\delta_{ij}$ is the Kronecker delta, and a fluctuating part $u'_i $ ($u'_i = u_i - \langle u_i \rangle_{xz}$). In this article we indicate the spatial average in the $x$ and $z$ directions with $\langle \cdot \rangle_{xz}$, fluctuations with the prime symbol ($'$), and the average in the full volume with $\langle \cdot \rangle$. The time evolution of the fluctuating velocity $u'_i$ is described by
\begin{align}
\label{eq:NS}
\rho \left( \frac{\partial u'_i}{\partial t} + \frac{\partial u'_i u'_j}{\partial x_j} + \mathcal{S} x_2 \frac{\partial u'_i}{\partial x_1} + \mathcal{S} u'_2 \delta_{i1} \right)  &= - \frac{\partial p}{\partial x_i} + \frac{\partial \tau_{ij}}{\partial x_j} + f_i, \\
\frac{\partial u'_i}{\partial x_i} &= 0,
\end{align}
where $\rho$ is the fluid density, $p$ is the pressure, $\tau_{ij}= 2 \mu \mathcal{D}_{ij}$ with $\mu$ the dynamic viscosity and $\mathcal{D}_{ij}$ the strain rate tensor ($\mathcal{D}_{ij}= \left( \partial u_i/ \partial x_j + \partial u_j/\partial x_i \right)/2$), and $f_i$ is the surface tension force defined as $f_i=\sigma \kappa n_i \delta$, where $\delta$ is the Dirac delta function at the interface, $\sigma$ the interfacial surface tension, $\kappa$ the interface curvature and $n_i$ the normal to the interface. This equation is written in the so-called one-fluid formulation \citep{tryggvason_sussman_hussaini_2007a} so that only one set of equations is solved in both phases. The problem is solved by introducing an indicator function $H$ to identify each fluid phase so that $H = 1$ in the region occupied by the suspended dispersed fluid (fluid $1$) and $H = 0$ in the carrier phase (fluid $2$). Considering that both fluids are transported by the flow velocity, we update $H$ in the Eulerian framework by the following advection equation written in divergence form,
\begin{equation}
\frac{\partial \phi}{\partial t} + \frac{\partial u_i H}{\partial x_i} = \phi \frac{\partial u_i}{\partial x_i},
\end{equation}
where $\phi$ is the cell-averaged value of the indicator function.
%Once $\phi$ is known, one can compute the fluid density and viscosity as
%\begin{equation}
%\rho = \rho_1 \phi + \rho_2 \left( 1- \phi \right) \;\;\;\;\; \textrm{and} \;\;\;\;\; \mu = \mu_1 \phi + \mu_2 \left( 1- \phi \right).
%\end{equation}

The above governing equations are solved numerically. First, the transport equation for $\phi$ is updated following the methodology described by \cite{ii_sugiyama_takeuchi_takagi_matsumoto_xiao_2012a} and \cite{rosti_de-vita_brandt_2019a} in order to obtain $\phi^{n+1}$ which is used to update the density and viscosity of the fluids. In particular, the indicator function $H$ is approximated as
\begin{equation} \label{eq:vofInd}
H \left( X, Y, Z \right) \approx \widehat{H} \left( X, Y, Z \right) = \frac{1}{2} \bigg( 1+ \tanh \big( \beta \left( P \left( X, Y, Z \right) + d \right) \big) \bigg),
\end{equation}
where $X, Y, Z \in \left[0, 1 \right]$ is a centered local coordinate system defined in each cell, $P$ is a three dimensional quadratic curved surface function determined algebraically by imposing the correct value of the three normal components and the six components of the Cartesian curvature tensor in each cell, $d$ is a normalization parameter used to enforce that the integral of the indicator function in each cell equals $\phi$ and $\beta$ is a sharpness parameter. $\beta$ is set equal to $1$ in the present work, the smallest value allowed by the method which ensures the sharpest possible interface for a given mesh size. Second, the momentum equation and the incompressibility constraint are solved following the method proposed by \cite{gerz_schumann_elghobashi_1989a} and recently adopted by \cite{tanaka_2017a}, in which the third term on the left-hand side of the momentum equation (\equref{eq:NS}), \ie the advection due to the mean shear flow, is solved separately using a Fourier approximation. In particular, the second-order Adams--Bashforth method is applied for the convection and viscous terms in \equref{eq:NS} to obtain an intermediate velocity
\begin{equation}
{u'_i}^*= {u'_i}^n + \Delta t \left( \frac{3}{2} \textrm{rhs}^n_i - \frac{1}{2} \textrm{rhs}^{n-1}_i \right),
\end{equation}
where $\Delta t$ is the time step from time $t^n$ to $t^{n+1}$ and
\begin{equation}
\textrm{rhs}_i=-\mathcal{S} u'_2 \delta_{i1} - \frac{\partial u'_i u'_j}{\partial x_j} + \frac{1}{\rho} \frac{\partial \tau_{ij}}{\partial x_j}.
\end{equation}
The time step $\Delta t$ is chosen such that the Courant--Friedrichs--Lewy (CFL) number $U_{max} \Delta t/\Delta x$ is smaller than unity, where $U_{max} = \mathcal{S} L_y$ is the maximum of the mean shear flow  velocity inside the computational domain. The advection due to the mean shear flow is then solved separately using a Fourier approximation as
\begin{equation}
{u'_i}^{**} \left( x_1, x_2, x_3 \right) = {u'_i}^* \left( x_1 - \Delta t \mathcal{S} x_2, x_2, x_3 \right).
\end{equation}
Note that \cite{tanaka_2017a} modified the approach of \cite{gerz_schumann_elghobashi_1989a} by performing a similar additional step for the pressure. Our tests suggest that the original form by \cite{gerz_schumann_elghobashi_1989a} is numerically more stable and physically consistent with the incompressibility constraint because the pressure is not a transported quantity. The surface tension term $f_i$ is then taken into account by updating the velocity field: we use the continuum surface force model by \cite{brackbill_kothe_zemach_1992a} to compute the surface tension force where the normals are obtained with the well known Youngs approach \citep{youngs_1982a}, \ie $f_i= \sigma \kappa \partial \phi/\partial x_i$, thus obtaining
\begin{equation}
{u'_i}^{***} = {u'_i}^{**} + \Delta t \frac{f_i^{n+1}}{\rho}.
\end{equation}
Then, we enforce the zero divergence of the velocity field by solving the following Poisson equation
\begin{equation}
\label{eq:poisson}
\frac{\partial^2 p^{n+1}}{\partial x_j \partial x_j} = \frac{\rho}{\Delta t} \frac{\partial {u'_i}^{***}}{\partial x_i},
\end{equation}
%Due to the non-uniformity of the density, the Poisson equation used to enforce a divergence-free velocity field results in an equation with variable coefficients, \ie
%\begin{equation}
%\label{eq:poisson}
%\frac{\partial}{\partial x_j} \left( \frac{1}{\rho^{n+1}} \frac{\partial p^{n+1}}{\partial x_j} \right)= \frac{1}{\Delta t} \frac{\partial {u'_i}^{***}}{\partial x_i}.
%\end{equation}
%In order to employ an efficient FFT-based pressure solver with constant coefficients \citep{dodd_ferrante_2014a, dodd_ferrante_2016a}, we use the following splitting of the pressure term \citep{dong_shen_2012a}:
%\begin{equation}
%\label{eq:pressureSplit}
%\frac{1}{\rho} \frac{\partial p}{\partial x_j} \to \frac{1}{\rho_0} \frac{\partial p}{\partial x_j} +\left( \frac{1}{\rho} - \frac{1}{\rho_0} \right)  \frac{\partial \widetilde{p}}{\partial x_j},
%\end{equation}
%where $\rho_0$ is a constant density equal to the lowest density of the two phases, and $\widetilde{p}$ is an approximated pressure obtained by linear extrapolation, \eg  $\widetilde{p} = 2 p^{n}-p^{n-1}$. With this splitting, the Poisson equation can be rewritten as
%\begin{equation}
%\label{eq:poissonSplit}
%\frac{\partial^2 p^{n+1}}{\partial x_j \partial x_j} = \frac{\rho_0}{\Delta t}\frac{\partial u^{***}_i}{\partial x_i} + \frac{\partial}{\partial x_i} \left[ \left( 1 - \frac{\rho_0}{\rho^{n+1}} \right) \frac{\partial \widetilde{p}}{\partial x_i}\right],
%\end{equation}
which is solved with a standard FFT-based solver by exploiting the periodic and shear-periodic boundary conditions as detailed in \cite{tanaka_2017a}. Finally, we correct the velocity with $p^{n+1}$ to enforce the incompressibility constraint
\begin{equation}
\label{eq:correcSplit}
{u'_i}^{n+1} = {u'_i}^{***} - \Delta t \frac{1}{\rho} \frac{\partial p^{n+1}}{\partial x_i}.
\end{equation}
Note that, our numerical scheme discretely conserves both momentum and kinetic energy (in absence of viscosity and surface tension) since we use second order centered finite difference on a staggered mesh and the divergence form of the convective terms \citep{morinishi_lund_vasilyev_moin_1998a}.
%\begin{equation}
%\label{eq:correcSplit}
%{u'_i}^{n+1} = {u'_i}^{***} - \Delta t \left[ \frac{1}{\rho_0} \frac{\partial p^{n+1}}{\partial x_i} + \left( \frac{1}{\rho^{n+1}} - \frac{1}{\rho_0} \right) \frac{\partial \widetilde{p}}{\partial x_i} \right].
%\end{equation}
%Note that, \equrefM{eq:poissonSplit} and \equrefMM{eq:correcSplit} reduce to the usual Poisson equation and correction step when the density is uniform, \ie when the density of the two fluids is the same.

\begin{table}
\centering
\setlength{\tabcolsep}{5pt}
\begin{tabular}{cccccccccc}
Case	&	Symbol	&	$D_0/L_z$	&	$\mathcal{N}_0$	&	$We_{\mathcal{S}_0}$	&	$We_{\textrm{rms}_0}$	&	$We_0$	&	$We_\lambda$	&	$Re_\lambda$	\\
\hline
$1$	&	-	&	$-$		&	$-$		&	$-$		&	$-$		&	$-$				&	$-$				&	$145$	\\
$2$	&	\textcolor{blue}{\large $\bullet$}	&	$0.36$	&	$4$		&	$0.2$	&	$0.04$	&	$0.5330$		&	$0.0220$		&	$83$	\\
$3$	&	\textcolor{green}{\large $\bullet$}	&	$0.16$	&	$51$		&	$0.2$	&	$0.2$	&	$0.8000$		&	$0.0776$		&	$101$	\\
$4$	&	\textcolor{brown}{\large $\bullet$}	&	$0.08$	&	$564$	&	$0.2$	&	$1$		&	$2.0943$		&	$0.9339$		&	$111$	\\
$5$	&	\textcolor{blue}{\tiny $\blacksquare$}	&	$0.36$	&	$4$		&	$1$		&	$0.2$	&	$2.0944$		&	$0.6773$		&	$113$	\\
$6$	&	\textcolor{green}{\tiny $\blacksquare$}	&	$0.16$	&	$51$		&	$1$		&	$1$		&	$4.0156$		&	$0.7536$		&	$117$	\\
$7$	&	\textcolor{brown}{\tiny $\blacksquare$}	&	$0.08$	&	$564$	&	$1$		&	$5$		&	$10.4717$		&	$4.9313$		&	$132$	\\
$8$	&	\textcolor{blue}{$\blacktriangle$}	&	$0.36$	&	$4$		&	$5$		&	$1$		&	$4.1890$		&	$2.0103$		&	$122$	\\
$9$	&	\textcolor{green}{$\blacktriangle$}	&	$0.16$	&	$51$		&	$5$		&	$5$		&	$7.9999$		&	$4.0868$		&	$131$	\\
$10$	&	\textcolor{brown}{$\blacktriangle$}	&	$0.08$	& 	$564$	&	$5$		&	$25$	&	$20.9432$	&	$13.3057$	&	$142$
\end{tabular}
\caption{Summary of the direct numerical simulations performed with different initial droplet sizes $D_0$, numbers of droplets $\mathcal{N}_0$ and surface tension $\sigma$, all at a fixed Reynolds number $Re_z = 15200$ and volume fraction $\Phi = 5\%$.}
\label{tab:cases}
\end{table}

\subsection{Setup}
The problem is governed by several dimensionless parameters, which define the problem under consideration. First, the computational box is defined by two aspect ratios $\AR_{xz} = L_x/L_z$ and $\AR_{yz} = L_y/L_z$ which are fixed equal to $2.05$ and $1.025$ respectively. These values have been chosen accordingly to what proposed by \cite{sekimoto_dong_jimenez_2016a} as ``acceptable" in the sense that they fall within the range of parameters in which the flow is as free as possible from box effects and can thus be used as a model of shear-driven turbulence in general. Indeed, homogeneous shear turbulence in an infinite domain evolves towards larger and larger length scales while simulations in a finite box are necessarily constrained to some degree by the box geometry. These authors noticed that the effect of the geometry can be reduced by ensuring that $L_z$ is the main constraint, thus resulting in the flow being ``minimal" in the spanwise direction. Next, once the size of the numerical box is fixed, to fully characterize the problem we define the shear Reynolds number based on the box width
\begin{equation}
Re_z = \frac{\mathcal{S} L_z^2}{\nu},
\end{equation}
the Weber number based on the initial droplet diameter $D_0$
\begin{equation}
\label{eq:wes}
We_{\mathcal{S}_0} = \frac{\rho \mathcal{S}^2 D_0^3}{\sigma},
\end{equation}
and the ratio of the initial droplet diameter to the box size $\AR_{Dz}=D_0/L_z$. In the following, we consider one case of single-phase flow as reference and nine cases of two-phase flows, all at the same Reynolds number equal to $15200$; in the multiphase cases, we vary the ratio $\AR_{Dz}$ and $We_{\mathcal{S}_0}$, as summarized in \tabref{tab:cases}. Note that, the Weber number here is mainly determined by the interfacial surface tension $\sigma$. Two other nondimensional parameters are the density and viscosity ratios, which are fixed to unity to study the individual effect of the Weber number (interfacial surface tension).

Besides the parameters just defined and based on the geometrical dimensions and initial and boundary conditions alone, in the following discussion we will use other nondimensional numbers because they turned out to be more relevant to understand the problem at hand; in particular, the two non-dimensional parameters which characterize the single-phase homogeneous shear turbulent flows, the Taylor-microscale Reynolds number $Re_\lambda$ and the shear-rate parameter $S^*$, defined as
\begin{equation}
\label{eq:rel}
Re_\lambda = \left( \frac{2\mathcal{K} }{3} \right)^{1/2} \frac{\lambda}{\nu} = \left( \frac{5}{3 \nu \varepsilon } \right)^{1/2} 2\mathcal{K},
\end{equation}
and
\begin{equation}
\label{eq:ss}
S^* = \frac{2 \mathcal{S} \mathcal{K} }{\varepsilon},
\end{equation}
where $\lambda = \sqrt{10\nu \mathcal{K}/\varepsilon}$ is the Taylor microscale \citep{sekimoto_dong_jimenez_2016a}, $\mathcal{K} = \langle \rho u'_i u'_i \rangle/2$ is the turbulent kinetic energy per unit volume, and $\varepsilon = \mu \langle \partial u'_i/\partial x_j \partial u'_i/\partial x_j \rangle$ is the dissipation rate of the fluctuating energy. These two non-dimensional numbers can be interpreted as the ratio of the eddy-turnover time $\tau_0= \left( 2\mathcal{K} \right)^{1/2}/\varepsilon$ and the Kolmogorov time scale $\tau_K= \left( \nu / \varepsilon \right)^{1/2}$ and the mean shear time scale $\tau_\mathcal{S} = 1/ \mathcal{S}$, respectively.

\begin{figure}
	\centering
	\input{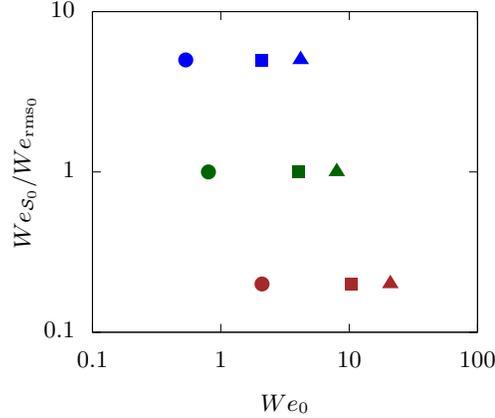} \vspace{0.5cm}
	\caption{The ratio of the two Weber numbers introduced here, one based on the mean shear $We_{\mathcal{S}_0}$ and one on the velocity fluctuations $We_{\textrm{rms}_0}$, as a function of the Weber number based on the initial droplet size, $We_0$. The circle, square and triangle symbols are used to distinguish cases with different surface tension but same ratio $We_{\mathcal{S}_0}/We_{\textrm{rms}_0}$, while the brown, green and blue colors represent cases with the ratio $We_{\mathcal{S}_0}/We_{{rms}_0}$ equal to $1/5$, $1$ and $5$, respectively. These symbols and color scheme will be used throughout the rest of the paper.}
	\label{fig:Web}
\end{figure}
Weber numbers can be defined in several ways. In \equref{eq:wes} we defined the Weber number based on the mean shear, but it can also be defined based on the velocity fluctuations, thus obtaining
\begin{equation}
\label{eq:werms}
We_{\textrm{rms}_0} = \frac{2 \rho \mathcal{K} D_0}{\sigma}.
\end{equation}
Note that, the latter definition is the one usually used in homogeneous isotropic turbulent flows in the absence of a mean flow \citep{dodd_ferrante_2016a}. Both the Weber numbers $We_{\mathcal{S}_0}$ and $We_{\textrm{rms}_0}$ are of interest since they are based on two different mechanisms that may affect the droplets dynamics: on large scales (large droplets) the effect of the mean shear is dominant, while on small scales (small droplets) the flow is mainly dominated by the isotropic turbulent fluctuations. Our set of parameters is chosen such that the ratio of these two Weber numbers $We_{\mathcal{S}_0}/We_{{rms}_0}$ equals $1/5$, $1$ and $5$, as reported in \figrefS{fig:Web}. In general both the mechanisms are present together and hence we can define a Weber number which incorporates both the effects as
\begin{equation}
\label{eq:we0}
We_0 = \frac{\rho \left( \sqrt{2\mathcal{K}} + \mathcal{S} D_0 \right)^2 D_0}{\sigma}.
\end{equation}
Finally, we can define a Weber number based on $\lambda$ as
\begin{equation}
\label{eq:wel}
We_\lambda = \frac{\rho \left( \sqrt{2\mathcal{K}} + \mathcal{S} \lambda \right)^2 \lambda}{\sigma}.
\end{equation}
The choice of using $\lambda$ in the definition of the Weber number instead of a dimension associated to the suspended phase is due to the fact that the interface is not only deforming, thus losing its original spherical shape, but also actively undergoing merging and break-up processes, which makes the definition of a unique dimension difficult. Therefore, we propose to rely on a fluid length scale, which, as shown below in the results, yields a good collapse of our data. In the following discussion, we use $We_\lambda$ to discuss the results; the value of $We_0$ is reported in order to fully characterize the initial conditions of the present simulations.

\subsection{Code validation}
The numerical code used in this work has been extensively validated in the past for multiphase turbulent flows simulations \citep{rosti_brandt_2017a, rosti_banaei_brandt_mazzino_2018a, rosti_izbassarov_tammisola_hormozi_brandt_2018a}. Here, we provide one more comparison with literature results for the specific case of HST.
\begin{figure}
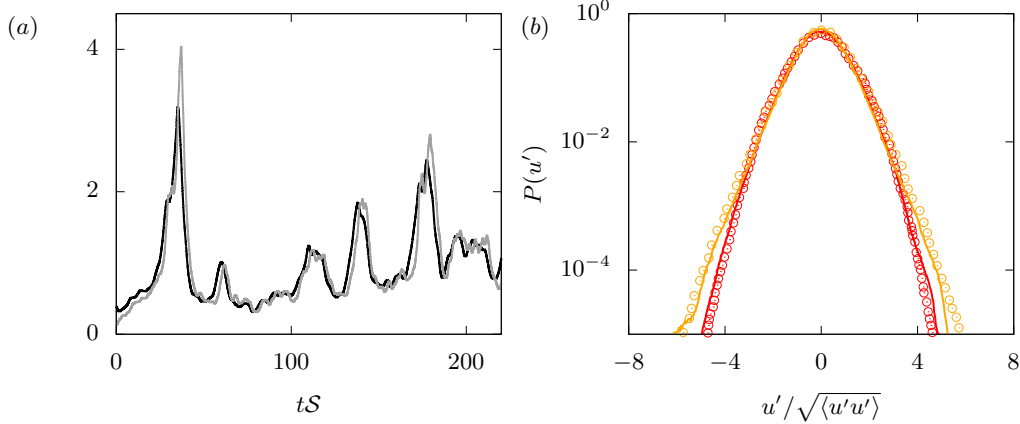

	\centering
	\input{pumir} \hspace{0.5cm}
	\input{pumirpdf} \vspace{0.5cm}
\caption{(a) Time history of the turbulent kinetic energy $\mathcal{K}=\langle u'_i u'_i \rangle/2$ (black line) and enstrophy $\Omega$ (grey line), normalized by their mean values. (b) Normalized histogram of the streamwise (red) and shear (orange) components of the velocity fluctuations, $u'$ and $v'$. The lines and symbols are used to distinguish our results (lines) from those by \citet{pumir_1996a}.}
	\label{fig:pumir}
\end{figure}
The single-phase homogeneous shear turbulence has been validated by reproducing one of the cases investigated by \citet{pumir_1996a}; in particular, we simulated the Run No.~$2$ in that paper. The initial condition at $t = 0$ is a homogeneous isotropic turbulent field at $Re_\lambda = 50.8$, obtained in a square computational box of size $2\pi$ discretised with $256$ grid points in each direction. From the time history of the turbulent kinetic energy $\mathcal{K}$ and of the enstrophy $\Omega = \langle \omega_i \omega_i \rangle$, shown in \figref[a]{fig:pumir}, we observe a first transient phase for $0 \le t\mathcal{S} \le 30$, where the kinetic energy and enstrophy grow rapidly, followed by a statistically stationary state characterized by a cyclic succession of turbulent kinetic energy peaks rapidly followed by a peak in enstrophy with a time lag of approximately $5\mathcal{S}$. This behavior is well captured in our simulation. A quantitative validation is performed first by comparing the mean components of the velocity anisotropy tensor, $b_{ij}=\langle u_i u_j / u'_k u'_k - \delta_{ij}/3 \rangle$ computed in our simulations ($b_{11}=0.231$, $b_{22} = −0.129$, $b_{33} = −0.101$, $b_{12} = −0.147$) with the data reported by \citet{pumir_1996a}, and we found that the differences are below $5\%$. A further comparison is shown in \figref[b]{fig:pumir} where the normalized histograms of the streamwise and shear components of the velocity obtained with the present  simulations are compared with the results reported in the literature \citep{pumir_1996a}; again we observe a very good agreement.

\section{Results} \label{sec:result}
\subsection{Statistically stationary state}
\begin{figure}
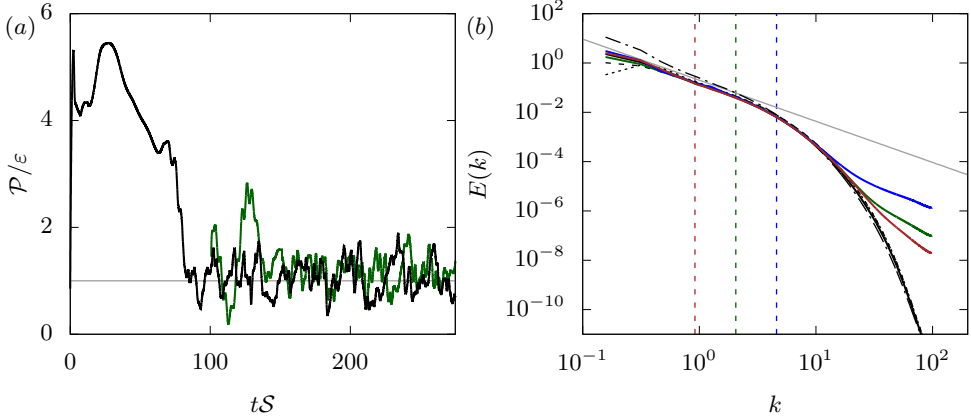

	\centering
	\input{tPE} \hspace{0.5cm}
	\input{kExyz} \vspace{0.5cm}
	\caption{(a) Time history of the ratio between the turbulent production $\mathcal{P} = - \langle u'v' \rangle d \langle u \rangle /dy$ and the turbulent dissipation rate $\varepsilon = \mu \langle \partial u'_i/\partial x_j \partial u'_i/\partial x_j \rangle$. The black and green lines represent the single and multiphase flows ($D_0=0.16L_z$ and $We_\lambda \approx 0.75$), respectively. (b) Spectra of the mean turbulent kinetic energy (black solid line) and its three spatial components (black dashed, dotted, and dashed-dotted lines) for the single-phase flow. The other three colored solid lines (blue, green and brown) are used for the spectra of the two-phase flows with $We_\lambda=0.02$, $0.75$ and $5$. The grey line is $\propto k^{-5/3}$, and the three vertical dashed lines represent the initial size of the droplets. The spectra are normalized by multiplying by $\varepsilon^{-2/3}$.}
	\label{fig:kin}
\end{figure}

We start our analysis by considering the single-phase flow at $Re_z = 15200$. The problem is solved numerically on a computational mesh of $1312 \times 640 \times 624$ grid points and the simulation is run for approximately $250\mathcal{S}$ time units. Note that, the grid spacing is chosen sufficiently small for good resolution of the smallest turbulent scales as indicated by $\Delta x / \eta \approx 0.7$, where $\eta$ is the Kolmogorov scale defined as $\eta=\left( \mu/\rho \right)^{3/4}/\varepsilon^{1/4}$. The initial flow field is  fully developed single-phase homogeneous isotropic turbulence, and the mean shear $\mathcal{S}$ is applied from $t=0$. As shown in \figref[a]{fig:kin}, once the shear is applied, the flow undergoes an initial transient characterized by a strong increase in the production of turbulent kinetic energy, which is not in balance with the dissipation rate. After some time, however, the turbulent kinetic energy $\mathcal{K}$ decreases owing to an increase in the dissipation, reaching a new statistically steady state where, on average, the production balances the dissipation ($\mathcal{P} \approx \varepsilon$). This state, called steady-state shear turbulence, was first found and characterized by \cite{pumir_1996a} and later investigated by others \citep[e.g.][]{sekimoto_dong_jimenez_2016a}. The resulting Taylor microscale Reynolds number at the steady state is equal to $Re_\lambda \approx 145$ with the averaged spectrum of the TKE reported in \figref[b]{fig:kin}. Owing to the high Reynolds number, a clear $k^{-5/3}$ regime develops at intermediate scales. We also observe that the spectra of each individual component of the velocity are different at small wave numbers because of the large-scale anisotropy, while all spectra coincide at higher wave numbers, consistently with what observed by \cite{pumir_1996a}.

We now consider the multiphase problem. After around $100 \mathcal{S}$, when the single-phase flow has already reached a statistically steady state, we inject spherical droplets into the domain at random locations, globally enclosing a volume fraction of the carrier phase of $5\%$. The initial droplet diameter $D_0$ is in the inertial range, as shown in \figref[b]{fig:kin} with the vertical dashed lines. In particular, three different initial diameters are chosen, $D_0/L_z \approx 0.08$ (brown), $0.16$ (green), and $0.32$ (blue), corresponding to approximately $1.1$, $2.5$, and $5.6$ times the single-phase Taylor microscale $\lambda$. After the introduction of the dispersed phase, a new short transient arises lasting approximately $50\mathcal{S}$, eventually leading to a new statistically steady state, as depicted in \figref[a]{fig:kin}. Also, in the multiphase case, we observe that, at regime, the turbulent production balances on average the dissipation rate ($\mathcal{P} \approx \varepsilon$).

The presence of the droplets modifies the flow profoundly. The averaged spectrum of the turbulent kinetic energy in both phases in the two-phase case is reported in \figref[b]{fig:kin}, where we observe that the interface mostly affects the large wave numbers (small scales) for which higher levels of energy are evident, while slightly lower energy is present at the large scales. Note that, the result is analogous to what was observed in decaying homogeneous isotropic turbulence for solid particles \citep{lucci_ferrante_elghobashi_2010a} and bubbles \citep{dodd_ferrante_2016a}; the increased energy at high wave numbers has been explained by the breakup of large eddies due to the presence of the suspended phase and the consequent creation of new eddies of smaller scale. In the same figure we can also observe that the effect of the droplets decreases as the Weber number increases; in other words, the spectra of the multiphase cases approach the single phase one as $We$ increases, while for low $We$ the spectra depart from the single phase case at smaller and smaller wavenumbers.

\begin{figure}
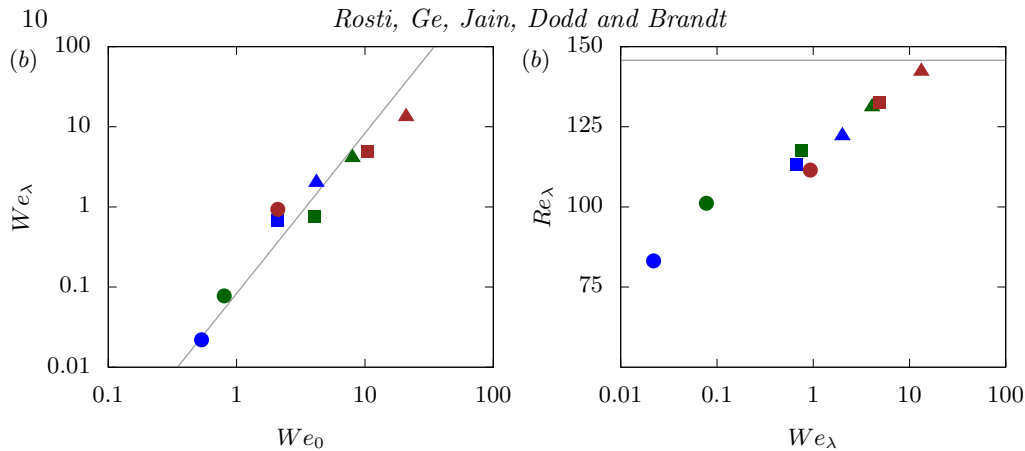

	\centering
	\input{We} \hspace{0.5cm}
	\input{Re_lambda2} \vspace{0.5cm}
	\caption{(a) Weber numbers based on the Taylor microscale, $We_\lambda$, as a function of the initial Weber number $We_0$ and (b) Reynolds numbers based on the Taylor microscale, $Re_\lambda$, as a function of $We_\lambda$. The grey solid line in the left panel is a fit to our data in the form of $We_\lambda \propto We_0^2$, while the grey solid line in the right panel represents the Taylor microscale Reynolds number $Re_\lambda$ of the single phase flow.}
	\label{fig:rey}
\end{figure}

As already discussed above, $We_0$ is the Weber number based on the initial droplet size, but since the droplets break up or coalesce, this measure is not fully representative of the final state of the multiphase problem; because of that, in the following sections we prefer to use the Weber number based on a flow length scale, $We_\lambda$, reported in \figref[a]{fig:rey} as a function of $We_0$. We can observe that the two Weber numbers are well correlated, with $We_\lambda$ scaling approximately as the square of $We_0$, \ie $We_\lambda \propto We_0^2$. The good level of correlation between the two definitions is a further demonstration that for the parameter range considered here the Weber number variations are mainly due to the changes of the interfacial surface tension rather than the chosen length scale.
\begin{figure}
	\centering
	\includegraphics[width=0.44\textwidth]{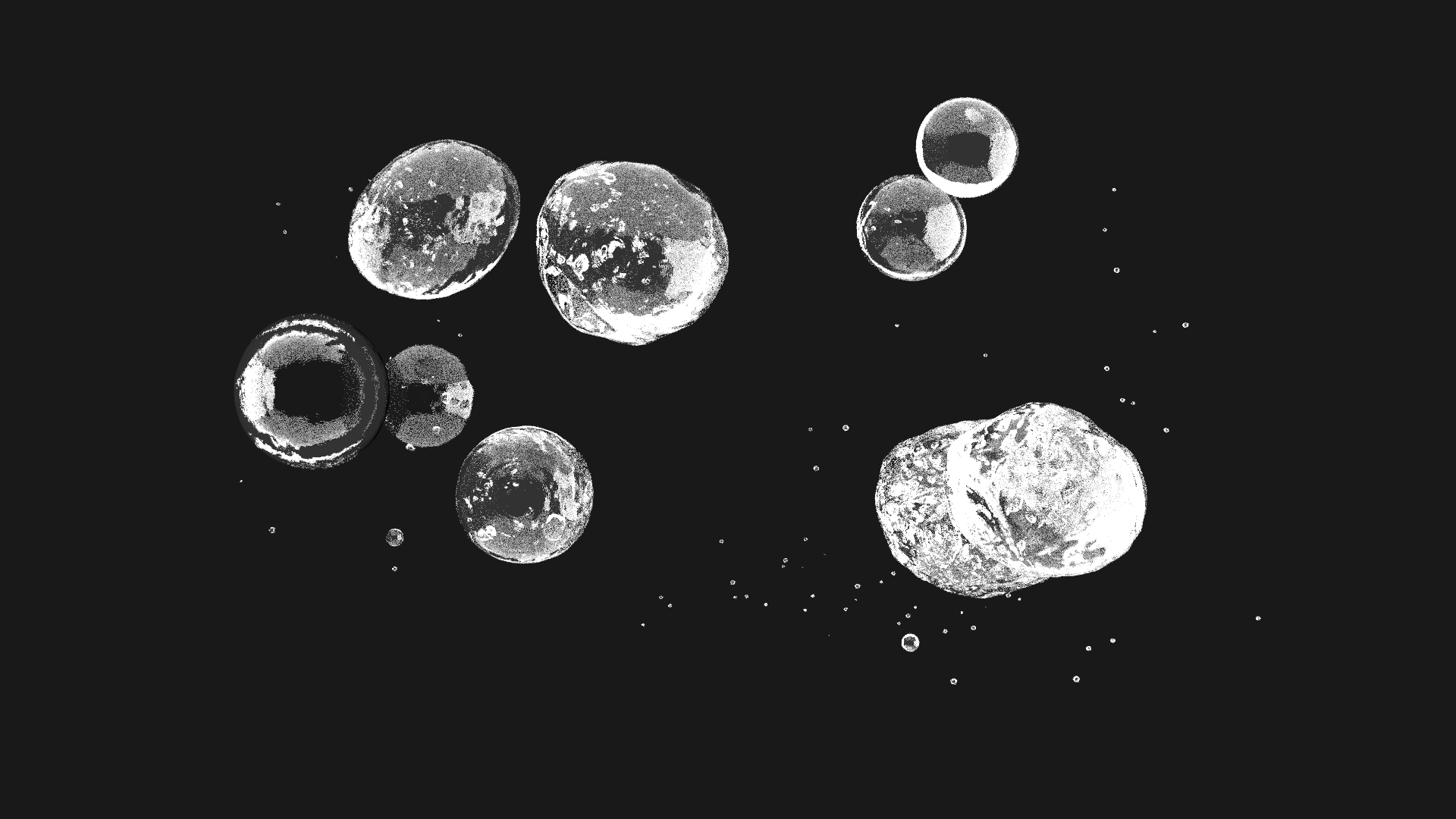}
	\includegraphics[width=0.44\textwidth]{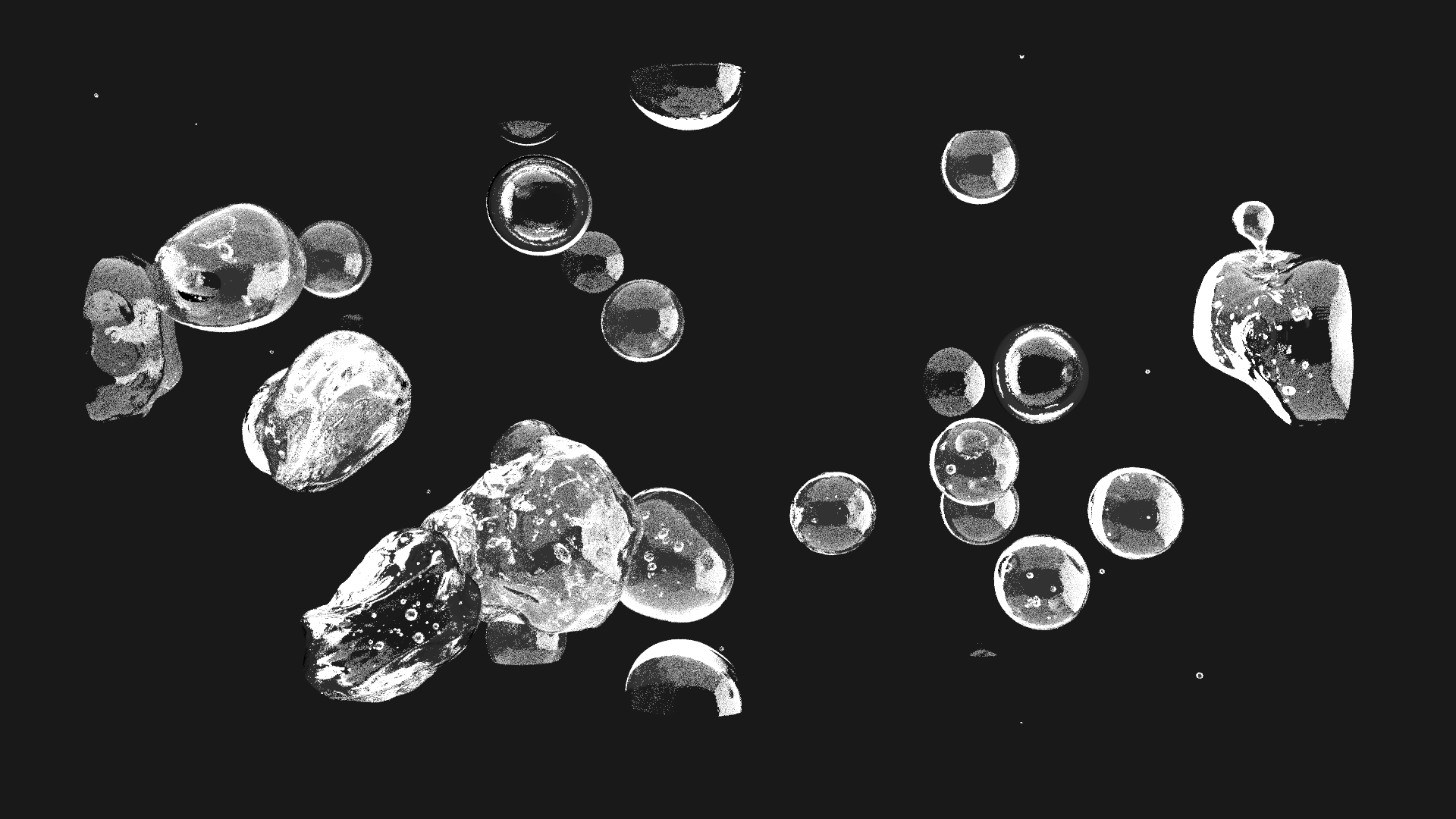} \\ \vspace{0.1cm}
	\includegraphics[width=0.44\textwidth]{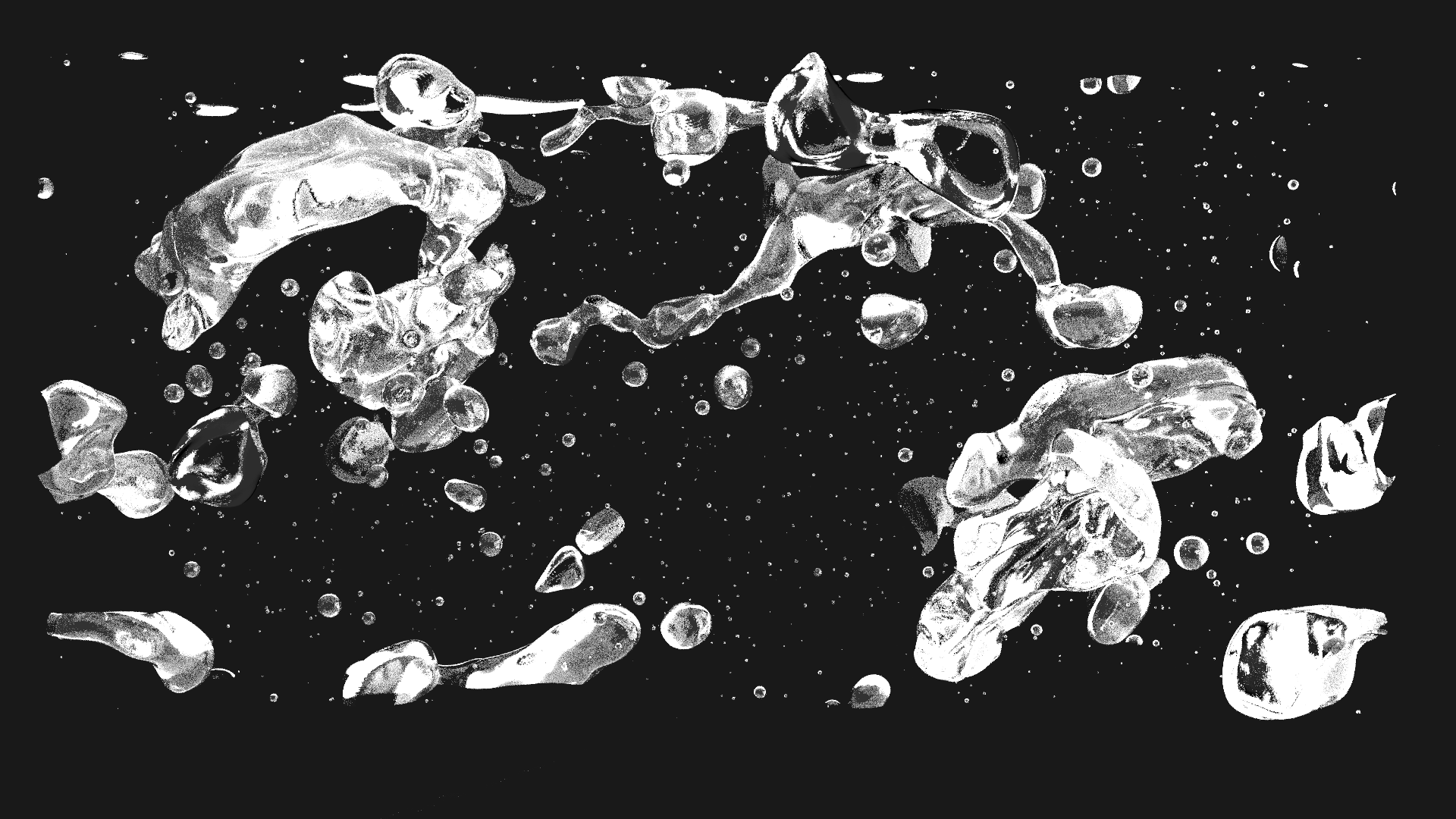}
	\includegraphics[width=0.44\textwidth]{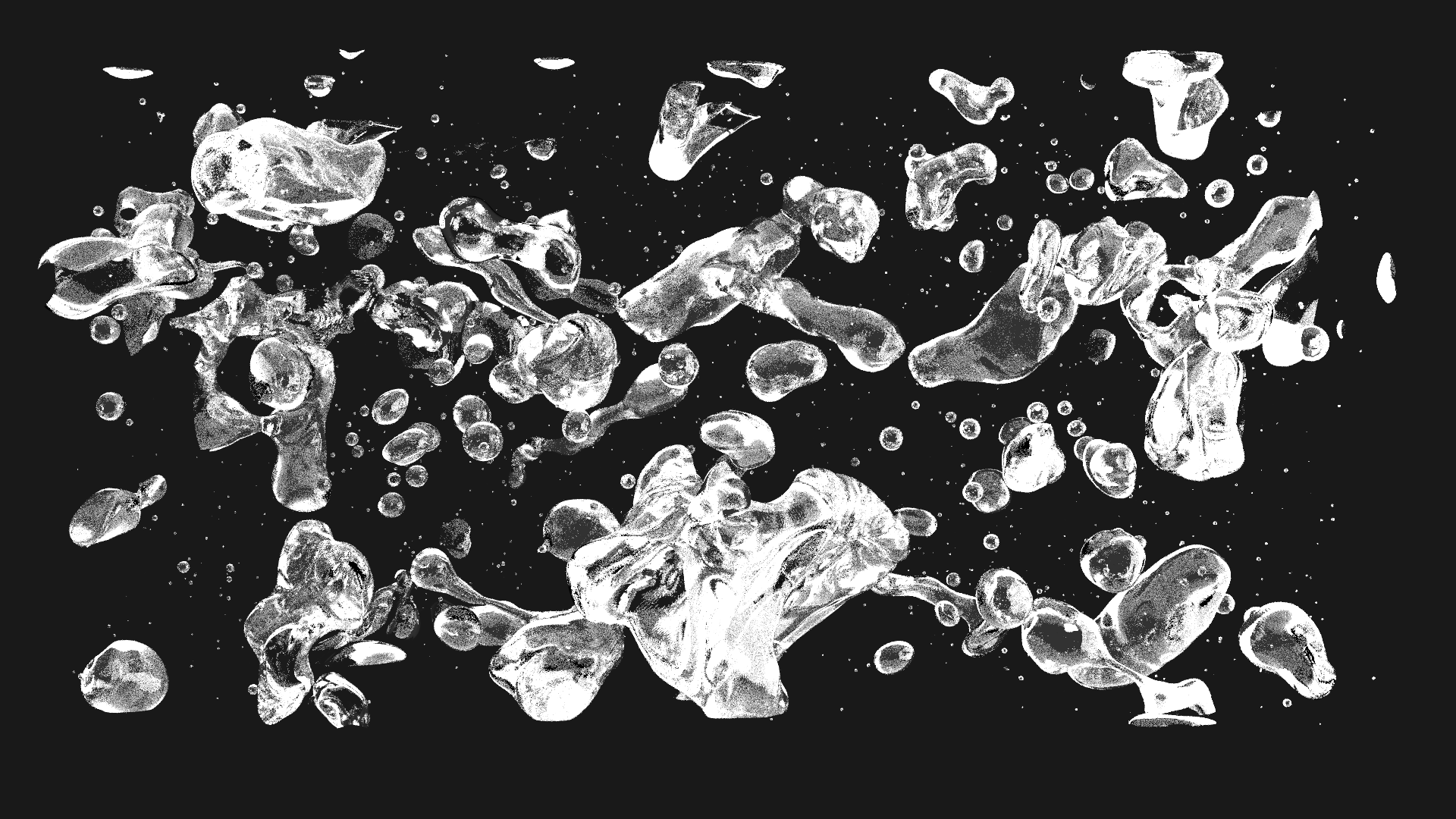} \\ \vspace{0.1cm}
	\includegraphics[width=0.44\textwidth]{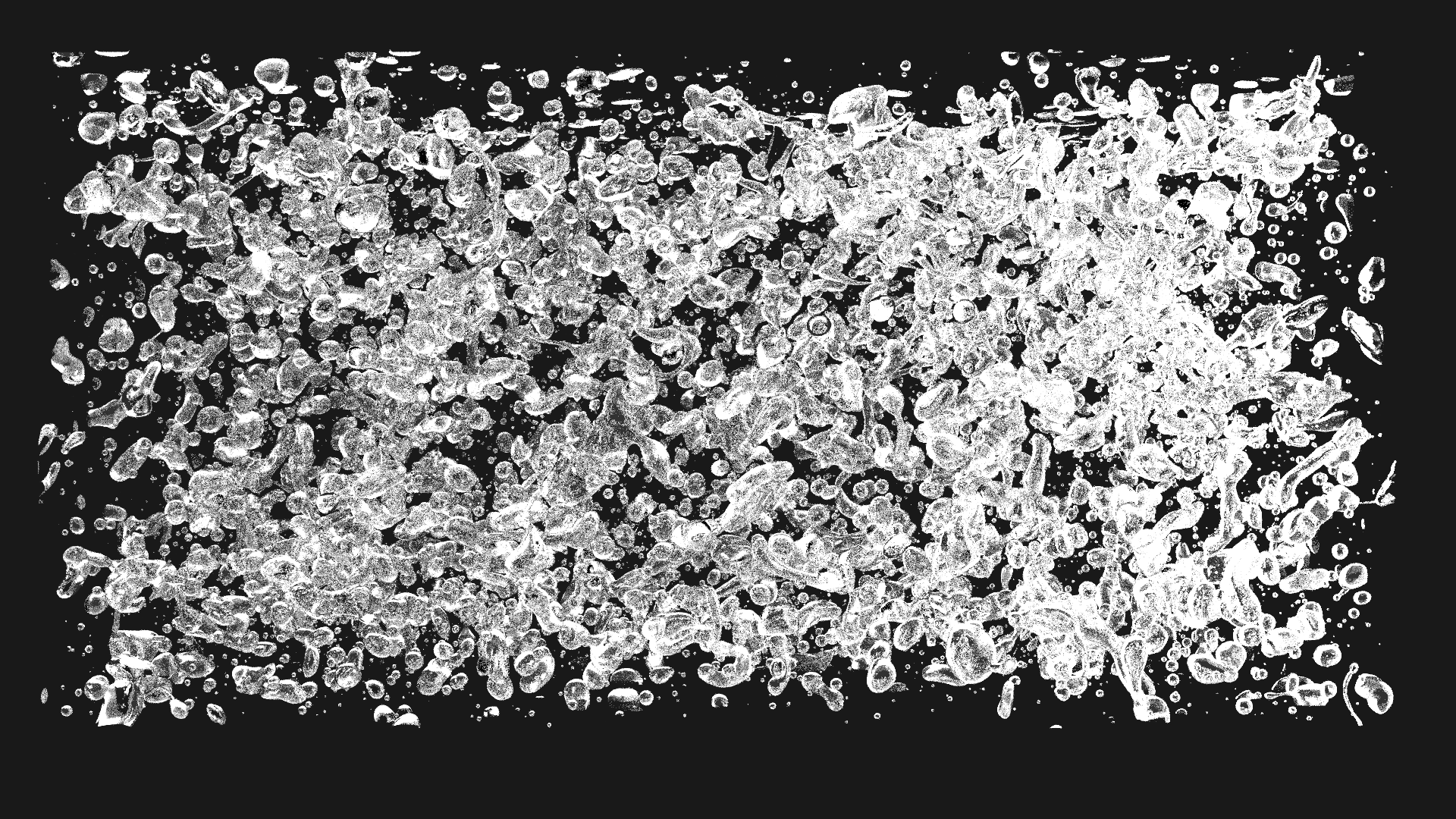}
	\includegraphics[width=0.44\textwidth]{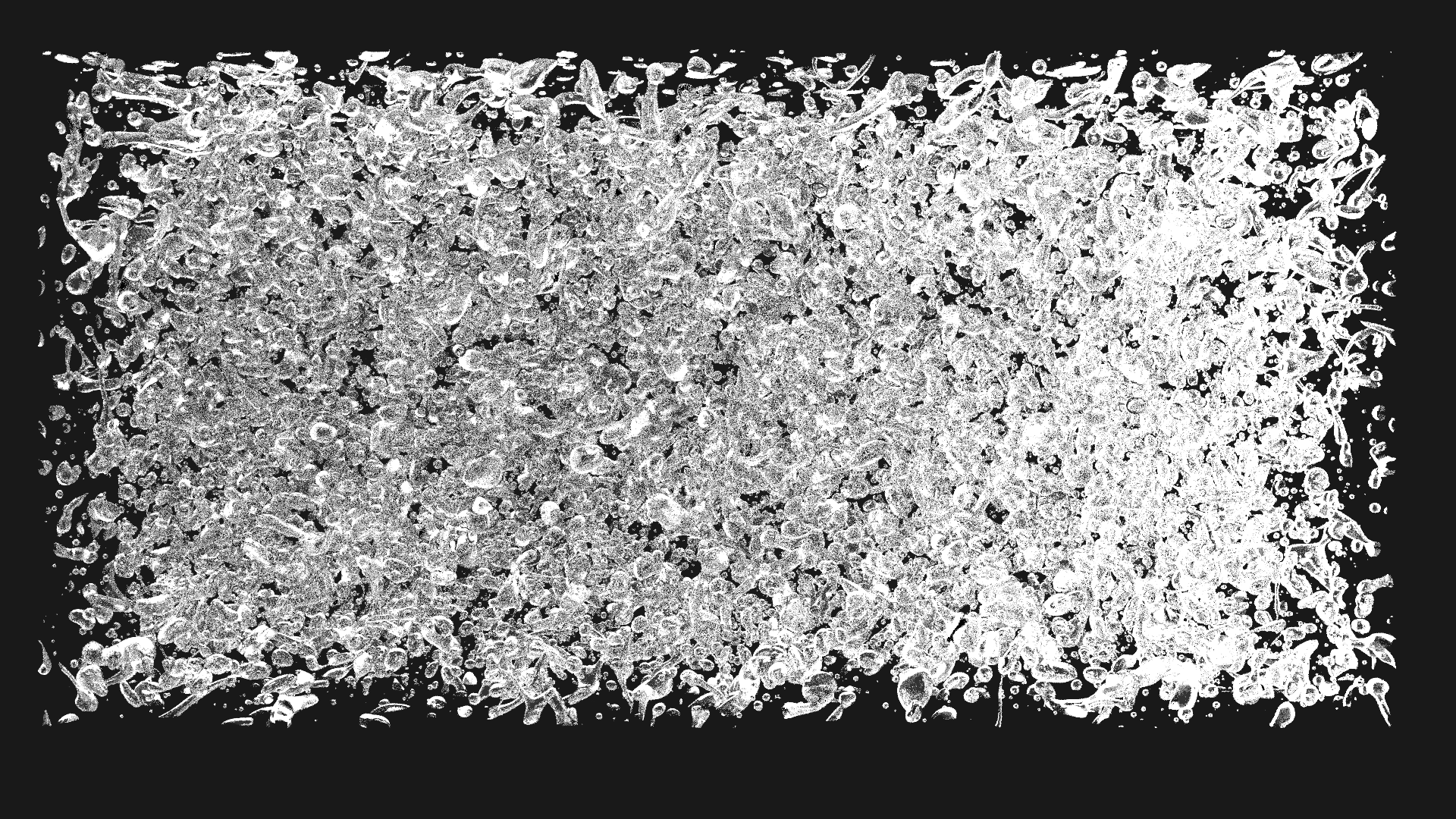}
	\caption{Visualisation in the $x-y$ plane of the interface in the homogeneous shear turbulent flow for different $We_\lambda$: (top left) $We_\lambda \approx 0.02$, (top right) $0.08$, (middle left) $0.8$, (middle right) $4$, (bottom left) $5$ and (bottom right) $13$. In the figures the flow is from left to right.}
	\label{fig:3dview}
\end{figure}

We quantify the turbulence modulation by examining the resulting $Re_\lambda$, shown for all our simulations in \figref[b]{fig:rey} as a function of the Weber number based on the Taylor microscale $We_\lambda$, and also reported in \tabref{tab:cases}. We can observe that the Reynolds number grows with $We_\lambda$ and that all the two-phase flow cases exhibit lower Taylor microscale Reynolds numbers than the single-phase case. Moreover, we observe that the difference decreases as the Weber number increases, with the two-phase flow cases approaching the single phase one as $We_\lambda$ increases, consistently with what was already observed in \figref[b]{fig:kin}. Indeed, the Reynolds number for the case with the most rigid droplets ($We_\lambda \approx 0.02$) is approximately half the single phase value ($-41\%$), while the difference with the single phase flow is only $2\%$ in the most deformable case ($We_\lambda \approx 13$). Note that, in the context of unbounded forced turbulent flows, such as homogeneous isotropic turbulence and homogeneous shear turbulence, a reduction of the Reynolds number can be interpreted as a drag increase, contrary to what is usually found in wall-bounded flows with constant flow rates where a reduction in the friction Reynolds number leads to drag decrease.

As first noteworthy result, the above data demonstrate that a statistical stationarity is not unique to single-phase homogeneous shear turbulent flows, but it is also realizable in the presence of a second, dispersed phase. Here, we have defined the stationary state in terms of the statistical properties of the flow averaged over both phases, but since the droplets can also break up or coalesce, it is natural to ask what the steady-state size distributions are and how that relates to the turbulence features. These questions are answered in the following sections.

\subsection{Size distribution}
We now study the transient and steady state property of the interface separating the two fluids. \figrefSC{fig:3dview} shows instantaneous snapshots of the two-phase flow at the statistically steady state, which is characterized by droplets with different sizes and shapes: in general we can observe that small droplets are approximately spherical, while the largest ones have very anisotropic shapes and show a preferential alignment with the direction of the mean shear. Also, as the Weber number decreases, the droplets size increases and larger droplets can sustain the spherical shape.

\begin{figure}
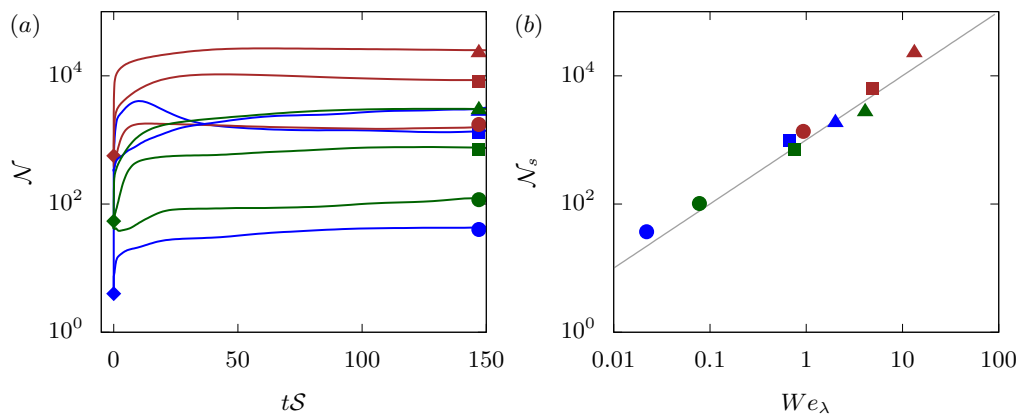

	\centering
	\input{Nt} \hspace{0.5cm}
	\input{N} \vspace{0.5cm}
	\caption{(a) Time history of the number of droplets $\mathcal{N}$ in the domain for different Weber numbers. The rombus symbols at $t=0$ represent the initial number of droplets. (b) The mean number of droplets $\mathcal{N}_s$ at the statistically steady state as a function of the Weber number $We_\lambda$. The grey solid line in the right panel is a fit to our data in the form of $\mathcal{N}_s \propto We_\lambda$.}
	\label{fig:count}
\end{figure}

\figrefC[a]{fig:count} shows the temporal evolution of the number of droplets ($\mathcal{N}$) under various $We_\lambda$ and initial sizes $D_0$. The counting of the droplets is conducted automatically by checking the connectivity of the local VOF field ($\phi$) using a $n$-dimensional image processing library\footnote{scipy.ndimage, https://docs.scipy.org/doc/scipy/reference/ndimage.html}. We observe that $\mathcal{N}$ has an initial transient phase of same duration as the fluid transient phase observed previously in \figref[a]{fig:kin} ($t\mathcal{S} \lesssim 50$), before the droplets count approaches a statistically steady value for all the cases considered, consistently with the statistically stationarity of the averaged global flow quantities. Note that, the final state is a statistically steady state since the number of droplets $\mathcal{N}$ is not constant but continuously varies and oscillates around a mean value, denoted later on as $\mathcal{N}_s$. From the figure we observe also that the initial transient phase differs among the cases, with three distinct behaviors evident: \textit{i)} in most of the cases, $\mathcal{N}$ increases rapidly after the injection (within $t \mathcal{S} \approx 10$); however, the growth slows down and $\mathcal{N}$ reaches its final steady state value almost monotonically; \textit{ii)} cases $4$ and $5$ exhibit a significant overshoot of the number of droplets $\mathcal{N}$ for short times before $\mathcal{N}$ reduces to the final regime values due to the coalescence; \textit{iii)} case $3$ shows an initial decrease of the number of droplets followed by an increase. Notwithstanding the different behaviors, in all the cases the final number of droplets is always larger than the initial one. 

The steady-state value of the number of droplets $\mathcal{N}_s$ as a function of $We_\lambda$ is reported in \figref[b]{fig:count}; we observe that $\mathcal{N}_s$ grows monotonically with $We_\lambda$ (see also the visualisations in \figrefS{fig:3dview}) and that the growth is nearly linear over the three decades spanned in the present study, \ie a fit to our data produces $\mathcal{N}_s\propto We_\lambda$ with an exponent of $1$. Since a high Weber number corresponds to a low surface energy, we conjecture that $\mathcal{N}_s$ grows indefinitely with $We_\lambda$. Note also that, cases $5$ and $6$ which have different initial droplet diameters, have almost the same final count of droplets $\mathcal{N}_s$ as well as $We_\lambda$. This provides additional evidence that the droplet statistics are better defined by the Weber number $We_\lambda$ based on the flow quantities rather than by that based on the initial droplet size $We_0$. These results suggest that the relative strength between the breakup and coalescence reflects the history of the flow features, and at equilibrium measurable quantities depend only on the global physical parameters.

\begin{figure}
	\centering
	\input{Cvol} \vspace{0.5cm} \\
\includegraphics[width=0.49\textwidth]{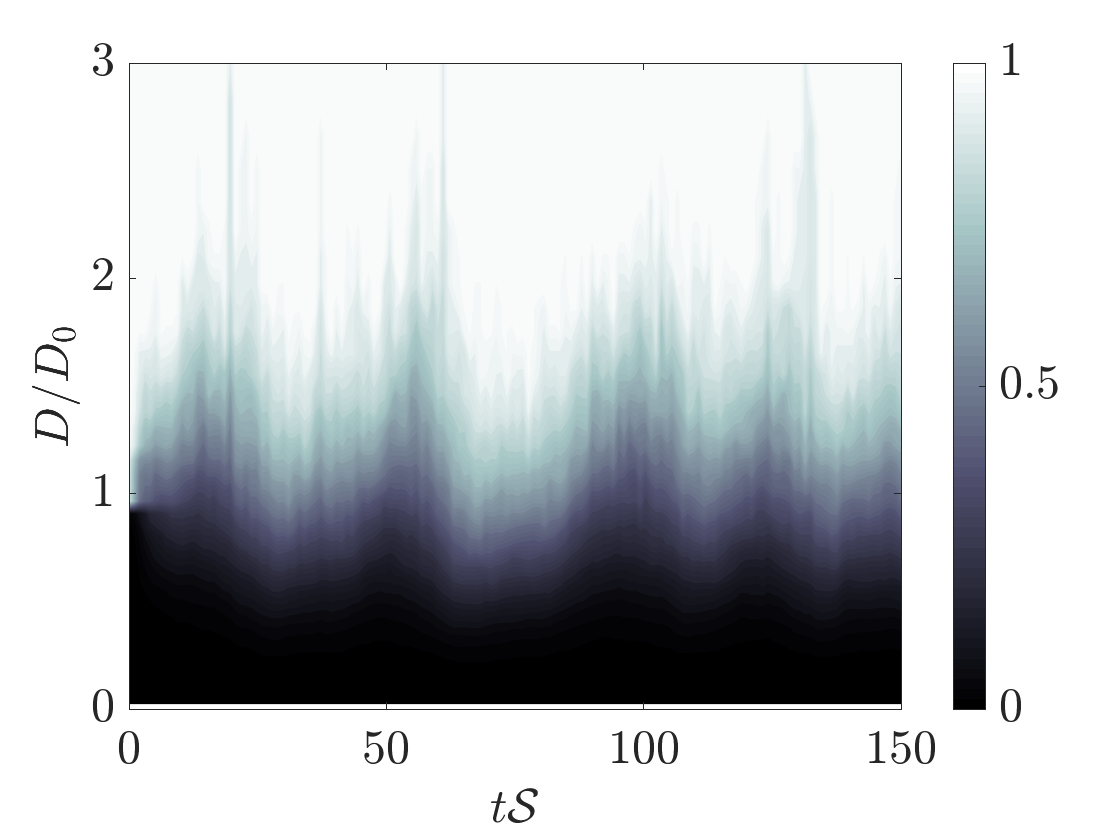}
\put(-190,130){$(b)$}
\includegraphics[width=0.49\textwidth]{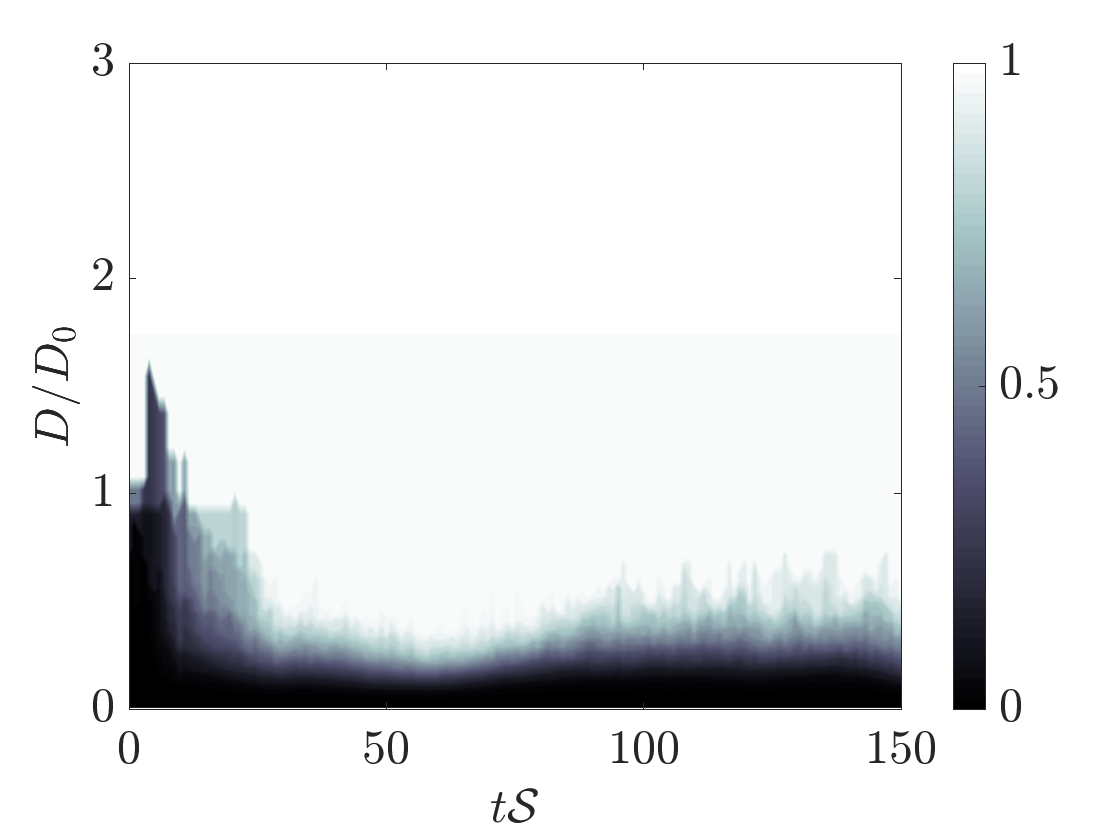}
\put(-190,130){$(c)$}
	\caption{(a) Normalized cumulative volume distributions $\mathcal{V}/\mathcal{V}_\textrm{tot}$ of the dispersed phase at the steady state as a function of the equivalent spherical droplet diameters $D$. The horizontal grey line correspond to the level $\mathcal{V}=0.95\mathcal{V}_\textrm{tot}$. (b-c) Contour of the temporal evolution of the normalized cumulative volume distributions of the dispersed phase as a function  of the equivalent spherical droplet diameter for cases $4$ (b) and $5$ (c).}
	\label{fig:distr}
\end{figure}
Next, we aim to characterize the steady state size distribution of the emulsion. Thus, we first examine the cumulative volume, $\mathcal{V}$, as a function of the equivalent spherical diameter $D$ defined as the diameter of the sphere occupying the same volume, see \figref[a]{fig:distr}. Specifically, \figref[a]{fig:distr} shows the steady-state distributions of all cases, where each point on the curves represents the total volume of the droplets with equivalent diameter lower than $D$. In the figure, both $\mathcal{V}$ and $D$ are normalized by the global maximal values so that the curves are bounded uniformly from above by $1$. The figure shows that the cumulative volume distribution only has one inflection point ($d^2 \mathcal{V}/dD^2 = 0$), thus indicating that the probability density plot ($d \mathcal{V}/dD$) is single peaked. In \figref[a]{fig:distr} the Weber number $We_\lambda$ grows from right to left, as indicated by the list of symbols, suggesting that small droplets tend to be more common at high Weber numbers. Additionally, the range of the droplet diameters also narrows with increasing $We_\lambda$, since the cumulative volume grows faster to unity, as visually confirmed in \figrefS{fig:3dview}. Case $2$, blue line with circle, is the only simulation exhibiting a double peak (\ie $d \mathcal{V}/dD$ has two local maxima): this is due to the presence of very small droplets together with few large ones as can be seen in \figref[left]{fig:3dview}. Nevertheless, the overall trend of decreasing size for increasing Weber number is still consistent with the linear scaling between $\mathcal{N}_s$ and $We_\lambda$, as already observed in \figref[b]{fig:count}. The two bottom panels in \figrefS{fig:distr}, are contours of $\mathcal{V}/\mathcal{V}_\textrm{tot}$ as a function of the equivalent diameter $D$ and time, and can thus be interpreted as a cumulative spectrogram with most of the droplets centered in the region where the gradient of the color is the largest. In particular, we selected two specific cases, with same initial Weber number $We_0 \approx 2$, but different initial droplet size and surface tension, thus leading to different $We_\lambda$. The two figures show the transient behavior for cases $4$ and $5$, respectively: in \figref[b]{fig:distr} the mean size distribution remains relatively unchanged over time but it is subject to strong fluctuations, while \figref[c]{fig:distr} shows a clear shift of the population from large droplets to small ones, with a statistically steady state characterized by small fluctuations.

\begin{figure}
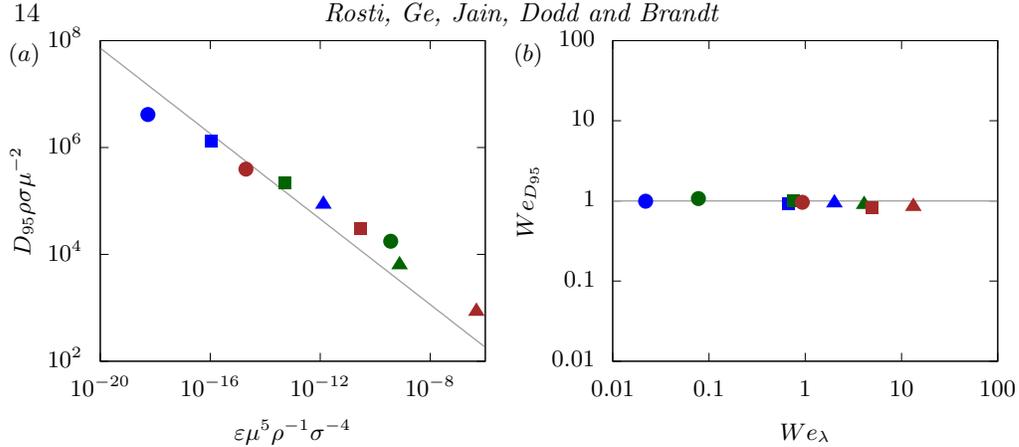

	\centering
	\input{hinze} \hspace{0.5cm}
	\input{WeH} \vspace{0.5cm}
	\caption{(a) Normalised maximum droplet size $D_{95}$ as a function of the energy input $\varepsilon$. The grey solid line is the relation $\rho \sigma D_{95}/\mu^2 = 0.725 \left( \mu^5 \epsilon/\rho \sigma^4 \right)^{-2/5}$ proposed by \cite{hinze_1955a}. (b) Critical Weber number $We_{D_{95}}$ based on the maximum droplet size $D_{95}$ for all the cases considered.}
	\label{fig:hinze}
\end{figure}

Another important parameter related to the size distribution is the largest droplet size, $D_\textrm{max}$. Assuming breakup of droplets due to the dynamic pressure ($\sim \rho U^2$), \cite{hinze_1955a} proposed that the largest possible droplet in a turbulent emulsifier is determined by the velocity fluctuation across $D_\textrm{max}$, \ie one can define a critical Weber number $We_\textrm{crit}=\rho \overline{u'^2} D_\textrm{max}/\sigma$, above which the droplet breaks up. \citet{hinze_1955a} showed that simple dimensional analysis leads to $D_\textrm{max} \propto \varepsilon^{-2/5}$, if isotropy prevails and the scaling by \citet{kolmogorov_1941a} is assumed valid. $D_\textrm{max}$ can be in general approximated by the diameter of the equivalent droplet occupying $95\%$ of the total dispersed volume, \ie $D_\textrm{max} \approx D_{95}$, which is represented in \figref[top]{fig:distr} with the dashed grey line. The symbols in the same figure provide the values of $D_{95}$ for our data. \figrefC[a]{fig:hinze} shows the normalised $D_{95}$ as a function of the scaled energy input, and indeed we can observe that our data scales with an approximately $-2/5$ slope. We remark that, although Hinze developed his theory considering only isotropic turbulent flows dominated by the breakup process and neglecting the coalescence, he hypothesized that the same scaling law might still hold for non-isotropic flows provided that the droplet sizes fall within the inertial range, such as in all our cases. More importantly, the success of the Hinze theory relies on the central assumption that breakup results from the dynamic pressure force, corresponding to a fixed critical Weber number. This is clearly shown in \figref[b]{fig:hinze}, which shows the Weber number based on $D_{95}$ as a function of $We_\lambda$. For all our cases, we obtain that $We_{crit} \approx 1$. Our results thus confirm that the $-2/5$ scaling between the maximum droplet diameter and the turbulence dissipation applies not only to isotropic turbulence, but also to the homogeneous shear turbulence that we have analyzed.

\begin{figure}
	\centering
	\input{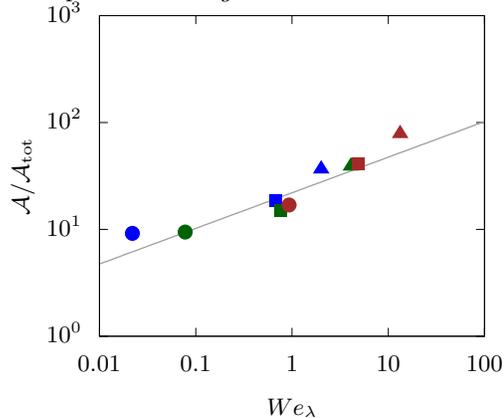} \vspace{0.5cm}
	\caption{Total interfacial area $\mathcal{A}$ as a function of the Weber number $We_\lambda$. The grey solid line is a fit to our data in the form of $\mathcal{A} \propto We_\lambda^{1/3}$.}
	\label{fig:surf}
\end{figure}
Finally, we can further characterize the size distribution of the emulsion by inspecting the total surface area $\mathcal{A}$ of the dispersed phase. This quantity is very important when studying multiphase flows with interfaces, since the rate of work due to the surface tension is equal to the product of the surface tension coefficient and the rate of change in interfacial surface area \citep{dodd_ferrante_2016a}; also, for many industrial applications, the total surface area is often the most important parameter as surfactants tend to reside on the interface or it determines the chemical reaction rate. \figrefSC{fig:surf} reports the steady state surface area $\mathcal{A}$ as a function of the Weber number $We_\lambda$ and clearly shows that the surface area increases monotonically with the Weber number. As we have shown above that $\mathcal{N} \propto We_\lambda$, combining with mass conservation, \ie $\mathcal{N} D^3 \propto 1$, leads to the following relation for the total area: $\mathcal{A} \propto \mathcal{N} D^2 \propto We_\lambda^{1/3}$. In other words, the surface area of the droplets shall also increase with the Weber number defined by the Taylor length of the flow, with a slope of $1/3$. \figrefSC{fig:surf} verifies this scaling. We remark that in the derivation above, we have assumed that the droplets are spherical, which is not always true in our cases. However, provided the linear scaling between $\mathcal{N}$ and $We_\lambda$ remains valid, we expect the $1/3$ scaling law to hold for a wide range of emulsions.

\subsection{Turbulent kinetic energy budget}
We now study how the multiphase nature of the problem affects the turbulent kinetic energy. To do so, we derive the turbulent kinetic energy evolution equation by first multiplying the momentum conservation equation \equref{eq:NS} by the velocity fluctuation $u'_i$,
\begin{equation}
\rho \left( \frac{\partial u'_i u'_i/2}{\partial t} + \frac{\partial u'_i u'_i u'_j/2}{\partial x_j} + \mathcal{S} x_2 \frac{\partial u'_i u'_i/2}{\partial x_1} + \mathcal{S} u'_1 u'_2 \right)  = - \frac{\partial u'_i p}{\partial x_i} + u'_i \frac{\partial  \tau_{ij}}{\partial x_j} + u'_i f_i.
\end{equation}
We make use of 
\begin{equation}
u'_i \frac{\partial \tau_{ij}}{\partial x_j} = \frac{\partial u'_i \tau_{ij}}{ \partial x_j} - \tau_{ij} \frac{\partial u'_i }{ \partial x_j} = \frac{\partial u'_i \tau_{ij}}{ \partial x_j} - \tau_{ij} \mathcal{D}_{ij},
\end{equation}
to obtain
\begin{equation}
\label{eq:tke0}
\rho \left( \frac{\partial u'_i u'_i/2}{\partial t} + \frac{\partial u'_i u'_i u'_j/2}{\partial x_j} + \mathcal{S} x_2 \frac{\partial u'_i u'_i/2}{\partial x_1} + \mathcal{S} u'_1 u'_2 \right)  = - \frac{\partial u'_i p}{\partial x_i} + \frac{\partial u'_i \tau_{ij}}{ \partial x_j} - \tau_{ij} \mathcal{D}_{ij} + u'_i f_i.
\end{equation}
\equrefC{eq:tke0} can then be either volume averaged over both phases to obtain the total kinetic energy equation, or phase averaged over the phase $m$ (\eg carrier or dispersed phase) to obtain the turbulent kinetic energy evolution equation for one phase only. 

The equation for the two-fluid mixture is obtained by applying the volume averaging operator
\begin{equation}
\langle \cdot \rangle = \frac{1}{\mathcal{V}} \int_\mathcal{V} \cdot \ \mathrm{d} \mathcal{V},
\end{equation}
leading to
\begin{equation}
\label{eq:tke_both}
\frac{d \mathcal{K}}{d t} = \mathcal{P} - \varepsilon + \Psi_\sigma,
\end{equation}
where the different terms indicate the rate of change of turbulent kinetic energy $\mathcal{K}$, the turbulent production rate $\mathcal{P}$, the dissipation rate $\varepsilon$ and the power of the surface tension $\psi_\sigma$, defined as
\begin{equation}
\mathcal{K} = \langle \rho u'_i u'_i \rangle/2, \ \
\mathcal{P} = -\mathcal{S} \langle \rho u'_1 u'_2 \rangle, \ \
\varepsilon = \langle \tau_{ij} \mathcal{D}_{ij} \rangle, \ \
\Psi_\sigma = \langle u'_i f_i \rangle.
\end{equation}
$\Psi_\sigma$ is the rate of work performed by the surface tension force on the surrounding fluid. It represents exchange of turbulent kinetic energy and interfacial surface energy and can be either positive or negative and thus a source or sink of turbulent kinetic energy. In particular, $\Psi_\sigma$ is proportional to the rate at which surface area is decreasing, \ie $\Psi_\sigma \propto -dA/dt$ \citep{dodd_ferrante_2016a}, and therefore decreasing (increasing) interfacial area through droplet restoration (deformation) or coalescence (breakup) is associated with $\Psi_\sigma$ being a source (sink) of turbulent kinetic energy. Note that all the transport terms in \equref{eq:tke0} vanish due to the homogeneity of the domain. On the other hand, if we apply the phase average operator
\begin{equation}
\langle \cdot \rangle_m = \frac{1}{\mathcal{V}_m} \int_{\mathcal{V}_m} \cdot \ \mathrm{d} \mathcal{V},
\end{equation}
we obtain
\begin{equation}
\label{eq:tke_single}
\frac{d \mathcal{K}_m}{d t} = \mathcal{P}_m - \varepsilon_m + \mathcal{T}^\nu_m + \mathcal{T}^p_m,
\end{equation}
where the different terms now indicate  the rate of change of turbulent kinetic energy $\mathcal{K}_m$, the turbulent production rate $\mathcal{P}_m$, the dissipation rate $\varepsilon_m$ and the viscous $\mathcal{T}^\nu_m$ and pressure $\mathcal{T}^p_m$ powers of the phase $m$, defined as
\begin{equation}\small
\mathcal{K} = \langle \rho u'_j u'_j \rangle_m/2, \ \
\mathcal{P}_m = -\mathcal{S} \langle \rho u'_1 u'_2 \rangle_m, \ \
\varepsilon = \langle \tau_{ij} \mathcal{D}_{ij} \rangle_m, \ \
\mathcal{T}^\nu_m = \left< \frac{\partial u'_i \tau_{ij}}{\partial x_j} \right>_m,  \ \
\mathcal{T}^p_m = - \left< \frac{\partial  u'_i p}{\partial x_i} \right>_m.
\end{equation}
In this case, the viscous and pressure transport terms are retained to account for a net flux of turbulent kinetic energy from one phase to the other caused by the coupling between the droplets and the carrier fluid (this physical interpretation can be seen more clearly by applying the Gauss's theorem to rewrite the terms as surface integrals, thus resulting in surface integration over the droplet surface). Note finally that the convective transport terms are zero because the two fluids are immiscible and therefore turbulent eddies can not transport turbulent kinetic energy across the interface.

\begin{figure}
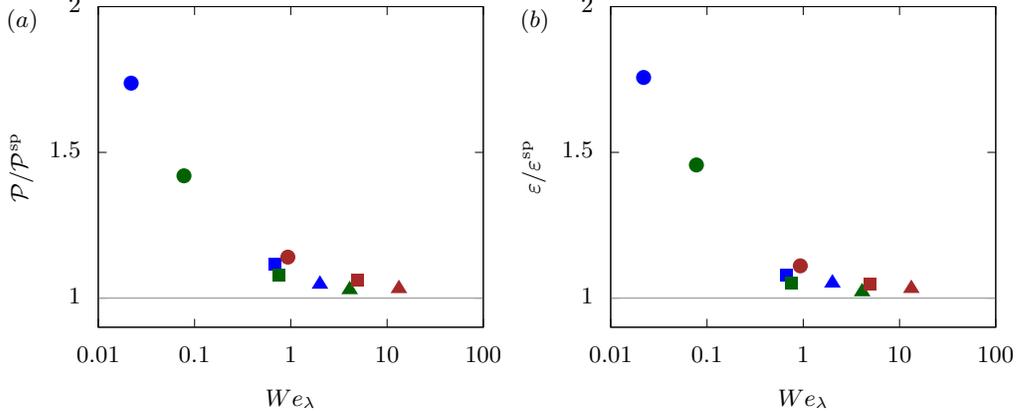

	\centering
	\input{P} \hspace{0.5cm}
	\input{eps} \vspace{0.5cm}
	\caption{(a) Turbulent kinetic energy production $\mathcal{P}$ and (b) dissipation $\varepsilon$ rates averaged over both phases as a function of the Weber number $We_\lambda$, normalized by their value in the single-phase flow ($\mathcal{P}_0$ and $\varepsilon_0$).}
	\label{fig:tke}
\end{figure}
First, we focus on the equation for $\mathcal{K}$ obtained by averaging over the whole volume and over both phases (\equref{eq:tke_both}). At steady state, the rate of change of $\mathcal{K}$ is obviously zero and the remaining terms are the production and dissipation rates and the power of surface tension. \figrefSC{fig:tke} shows the production $\mathcal{P}$ and dissipation $\varepsilon$ rates, normalized by their single-phase values $\mathcal{P}_0$ and $\varepsilon_0$, for all the simulations performed in the present study as a function of the Weber number $We_\lambda$. We observe that both the normalized production and dissipation rates are greater than unity and decrease monotonically as the $We_\lambda$ increases, indicating that the presence of the droplets leads to turbulence augmentation. As $We_\lambda$ decreases, the droplets become increasingly rigid, and therefore they exert a blocking effect on the surrounding turbulent flow. This effect abruptly re-orients the turbulent eddies leading to an increase in the magnitude of the Reynolds stress, $\langle u_1' u_2' \rangle$, causing an increase in $\mathcal{P}$, which also leads to an increase in the magnitude of the velocity gradients $\mathcal{D}_{ij}$, associated with an increase in $\varepsilon$ relative to the single-phase flow, as shown in \figrefS{fig:tke}. Moreover, the two quantities have approximately the same value (the difference is less than $3\%$), thus indicating that at steady state the production balances the dissipation and that the power of surface tension is on average zero (\ie $\mathcal{P} \approx \varepsilon$ and $\Psi_\sigma \approx 0$). These results are consistent with what was previously observed in \figref[a]{fig:kin} and indirectly confirm the relation $\Psi_\sigma= - \sigma/\mathcal{V}_m~d\mathcal{A}/dt$ derived by \citet{dodd_ferrante_2016a}. Indeed, this relation implies that at steady state $\Psi_\sigma$ is zero since the rate of change of $\mathcal{A}$ is null.

\begin{figure}
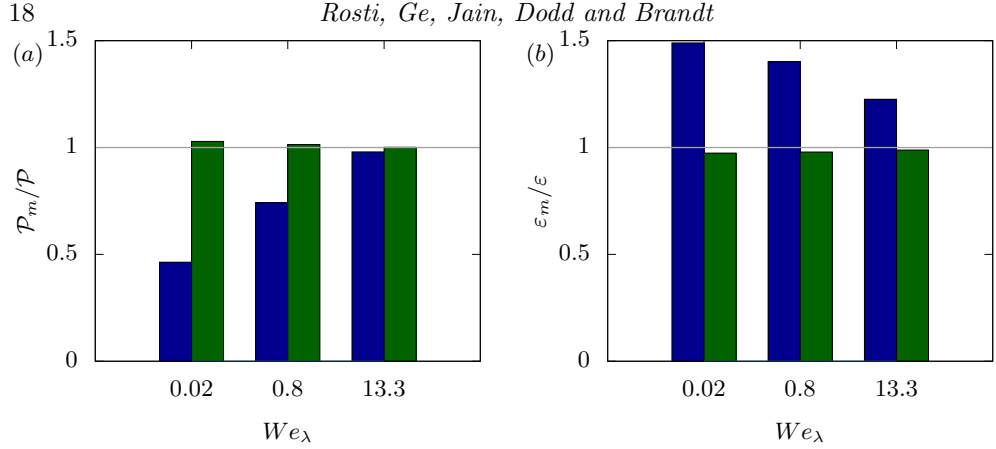

	\centering
	\input{histProd} \hspace{0.5cm}
	\input{histEps} \vspace{0.5cm}
	\caption{(a) Turbulent kinetic energy production $\mathcal{P}_m$ and (b) dissipation $\varepsilon_m$ rates averaged over the two phases separately as a function of the Weber number $We_\lambda$ for cases $2$, $6$ and $10$. The left and right columns are used to distinguish the dispersed and carrier phases, respectively.}
	\label{fig:tkem}
\end{figure}
Next, we focus on the equation obtained by phase averaging in one of the two fluids (\equref{eq:tke_single}). Again, at steady state the time derivative on the left-hand side is zero and the relation states that the production and dissipation are balanced by the two transport terms $\mathcal{T}^\nu_m$ and $\mathcal{T}^p_m$. \figrefSC{fig:tkem} shows histograms of the production $\mathcal{P}_m$ and dissipation $\varepsilon_m$ rates in the two phases for three selected Weber numbers $We_\lambda$ (cases $2$, $6$ and $10$). We observe that the production rate is lower in the dispersed phase than in the carrier phase, while the dissipation rate is higher in the dispersed fluid than in the carrier fluid. These results indicate that the total transport term $\mathcal{T}_m=\mathcal{T}^\nu_m + \mathcal{T}^p_m$ is positive in the dispersed fluid and negative in the carrier, corresponding to a turbulent kinetic energy transfer from the carrier to the dispersed phase. In other words, the presence of the droplets is overall a sink for the turbulent kinetic energy of the bulk fluid $\mathcal{K}_c$. In addition, we observe that the difference in $\mathcal{P}_m$ and $\varepsilon_m$ decreases with $We_\lambda$.

\begin{figure}
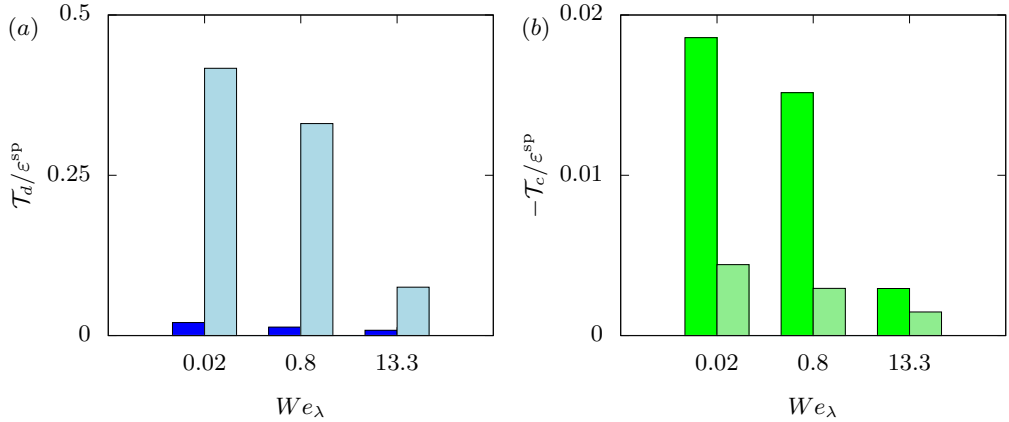

	\centering
	\input{histTranD} \hspace{0.5cm}
	\input{histTranC} \vspace{0.5cm}
	\caption{(a) Dispersed and (b) carrier transport terms $\mathcal{T}_m$, averaged over the two phases separately as a function of the Weber number $We_\lambda$ for cases $2$, $6$ and $10$. The left and right columns are used to distinguish the pressure and viscous contributions, respectively.}
	\label{fig:tkem_tran}
\end{figure}
Finally, \figrefS{fig:tkem_tran} shows the decomposition of the total transport term $\mathcal{T}_m$ into its pressure and viscous contributions, $\mathcal{T}^p_m$ and $\mathcal{T}^\nu_m$. In the dispersed phase shown in the left panel, the pressure transport term is very small and almost negligible, with most of the transport of turbulent kinetic energy ($90$--$95\%$) due to the viscous contribution $\mathcal{T}^\nu_d$. On the other hand, an opposite behavior is evident in the carrier phase shown in the right panel: the pressure transport term $\mathcal{T}^p_c$ is dominant one and accounting for most of the transport of turbulent kinetic energy ($65$--$80\%$), while the pressure contribution is small. Moreover, we can observe that all the transport terms reduce for increasing Weber number, consistently with the discussion concerning \figrefS{fig:tkem}.

\begin{figure}
	\centering
	\input{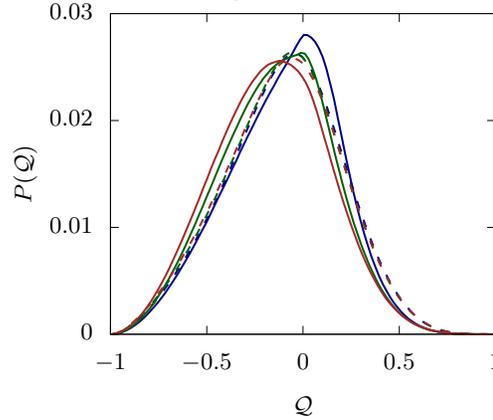} \vspace{0.5cm}
	\caption{Probability density function of the flow topology parameter $\mathcal{Q}$ for three different Weber numbers: cases $2$ (blue line), $6$ (green line) and $10$ (brown line), same as \figrefS{fig:tkem_tran}. The solid and dashed lines are used for the dispersed and carrier phase, respectively.}
	\label{fig:flowtop}
\end{figure}
The different mechanism of transport of turbulent kinetic energy between the carrier and dispersed phase is due to the different kind of flow experienced by the two fluids. This is discussed in \figrefS{fig:flowtop} where the so-called flow topology parameter $\mathcal{Q}$ \citep[see e.g.][]{de-vita_rosti_izbassarov_duffo_tammisola_hormozi_brandt_2018a} is presented. The flow topology parameter is defined as
\begin{equation}
\label{eq:flowtop}
\mathcal{Q} = \frac{\mathcal{D}^2 - \Omega ^2}{\mathcal{D}^2 + \Omega^2},
\end{equation}
where $\mathcal{D}^2 = \mathcal{D}_{ij}\mathcal{D}_{ji}$ and $\Omega^2 = \Omega_{ij}\Omega_{ji}$, being $\Omega_{ij}$ the rate of rotation tensor, $\Omega_{ij} = (\partial u_i/\partial x_j - \partial u_j/\partial x_i)/2$. When $\mathcal{Q} = -1$ the flow is purely rotational, regions with $\mathcal{Q} = 0$ represent pure shear flow and those with $\mathcal{Q} = 1$ elongational flow. The distribution of the flow topology parameter for three selected cases is reported in \figrefS{fig:flowtop}. Note that, in the figure we show the probability density function (pdf) of $\mathcal{Q}$ in the two liquid phases separately. We observe that in the carrier fluid (dashed lines) the flow is mostly a shear flow as demonstrated by a single broad peak at $\mathcal{Q} = 0$, and that little changes when changing the Weber number. On the other hand, the flow of the dispersed fluid (solid lines) still shows a broad single peak, now shifted towards negative values of $\mathcal{Q}$, meaning that the flow is more rotational. Also, the relevance of the rotational flow is more and more evident as the Weber number increases. This is caused by the increased number of droplets and their consequent reduction in size: indeed, as the droplets size reduces the effect of the shear reduces as well.

\section{Conclusions} \label{sec:conclusion}
We perform direct numerical simulations of two-phase homogeneous shear turbulent flows at $Re_z = 15200$, where the two-phase nature of problem is tackled numerically using the MTHINC volume of fluid method recently developed. The droplets are initially spheres providing $5\%$ volume fraction of the suspended phase and various Weber numbers and droplet initial diameters are investigated.

We show that the two-phase flow is able to reach a statistically steady state as indicated by a balance of turbulent kinetic energy production and dissipation. The results show that the presence of the droplets leads to turbulence augmentation by increasing the dissipation and production rates of the turbulence relative to the droplet-free flow. In particular, we find that as the Weber number decreases (higher droplet surface tension), the dissipation rate increases, causing the Taylor-microscale Reynolds number to decrease. This is explained by the surface tension force exerting a blocking effect on the surrounding turbulent flow. The turbulent production and dissipation rates are on average equal and in balance, with values larger than their single phase counterparts. Also, the surface tension power is on average zero. The flow modifications are caused by the presence of the dispersed phase, which acts as a sink of turbulent kinetic energy for the carrier phase, with a net flux going from the bulk of the fluid to the dispersed phase where it is dissipated. Moreover, the transport of turbulent kinetic energy in the carrier fluid is mainly due to the pressure transport, while the one inside the dispersed phase is dominated by the viscous contribution. This difference is explained by the different nature of the flow in the two phases: the carrier fluid is mainly a shear flow, while the dispersed fluid is more rotational owing to its smaller length scales where the effect of the mean shear is reduced.

In addition to the flow properties, the droplet distribution eventually reaches a statistically stationary condition. Indeed, we show that the flow reaches a condition where the number of droplets remains almost constant, due to a balance between the break up and coalescence mechanisms, and that the number of droplets grows approximately linearly with the Weber number. A similar trend is found for the averaged surface area which also grows monotonically with the Weber number, but the growth rate is less than linear (the surface area grows with the Weber number to the power of $1/3$, at least for moderately large Weber numbers). With the exception of one case, the droplet size distribution is single peaked, with the mean droplet size reducing with the Weber number. Based on the size distribution data, we show that the maximum droplets size scales well with the energy input as proposed by \citet{hinze_1955a}, although the possibility of coalescence mechanism and the presence of a mean shear which were not considered in the original formulation by \citet{hinze_1955a}.

\section*{Acknowledgments}
M.R.\ and L.B.\ acknowledge financial support by the European Research Council Grant no. ERC-2013-CoG-616186, TRITOS and by the Swedish Research Council Grant no. VR 2014-5001. S.S.J.\ was supported by the Franklin P.\ and Caroline M.\ Johnson Fellowship. The authors acknowledge computer time provided by the Swedish National Infrastructure for Computing (SNIC), the National Infrastructure for High Performance Computing and Data Storage in Norway (project no. NN9561K), and the Certainty cluster awarded by the National Science Foundation to CTR. Finally, special thanks are given to the CTR for hosting M.R.\, Z.G.\ and L.B.\ in the 2018 CTR Summer Program and to the other participants of the Multiphase Flows group for useful discussions.

%M.R.\ and L.B.\ were supported by the European Research Council Grant no. ERC-2013-CoG-616186, TRITOS and by the Swedish Research Council Grant no. VR 2014-5001. S.S.J.\ was supported by the Franklin P.\ and Caroline M.\ Johnson Fellowship. The authors acknowledge computer time provided by the Swedish National Infrastructure for Computing and by the National Infrastructure for High Performance Computing and Data Storage in Norway (project no. NN9561K). The authors also acknowledge use of computational resources from the Certainty cluster awarded by the National Science Foundation to CTR.

\bibliographystyle{jfm}
\bibliography{../../../../../../Articles/bibliography.bib}

\end{document}